\documentclass[reprint,superscriptaddress,amssymb,amsmath,aps,prapplied,longbibliography]{revtex4-1}

\usepackage{graphicx}
\usepackage{amssymb}
\usepackage{epstopdf}
\usepackage{upgreek}
\usepackage{dcolumn}
\usepackage{bm}
\usepackage{gensymb}
\usepackage{braket}
\usepackage{amsmath}
\usepackage{mathtools}
\usepackage{float}
\usepackage{tikz}

\expandafter\ifx\csname package@font\endcsname\relax\else
\expandafter\expandafter
\expandafter\usepackage
\expandafter\expandafter
\expandafter{\csname package@font\endcsname}%
\fi

\newcommand{\microamp}{$\upmu$A}
\newcommand{\micron}{$\upmu$m~}

\newcommand{\be}{\begin{eqnarray}}
\newcommand{\ee}{\end{eqnarray}}
\newcommand{\bfig}{\begin{figure}}
	\newcommand{\efig}{\end{figure}}

\DeclareFontFamily{U}{mathb}{}
\DeclareFontShape{U}{mathb}{m}{n}{
	<-5.5> mathb5
	<5.5-6.5> mathb6
	<6.5-7.5> mathb7
	<7.5-8.5> mathb8
	<8.5-9.5> mathb9
	<9.5-11.5> mathb10
	<11.5-> mathbb12
}{}

\usepackage[colorlinks=true,allcolors=blue]{hyperref}%

\begin{document}
	
	\title{A three-wave mixing kinetic inductance traveling-wave amplifier with near-quantum-limited noise performance}
	\author{M. Malnou}
	\email{maxime.malnou@nist.gov}
	\affiliation{National Institute of Standards and Technology, Boulder, Colorado 80305, USA}
	\affiliation{Department of Physics, University of Colorado, Boulder, Colorado 80309, USA}
	\author{M. R. Vissers}
	\affiliation{National Institute of Standards and Technology, Boulder, Colorado 80305, USA}
	\author{J. D. Wheeler}
	\affiliation{National Institute of Standards and Technology, Boulder, Colorado 80305, USA}
	\author{J. Aumentado}
	\affiliation{National Institute of Standards and Technology, Boulder, Colorado 80305, USA}
	\author{J. Hubmayr}
	\affiliation{National Institute of Standards and Technology, Boulder, Colorado 80305, USA}
	\author{J. N. Ullom}
	\affiliation{National Institute of Standards and Technology, Boulder, Colorado 80305, USA}
	\affiliation{Department of Physics, University of Colorado, Boulder, Colorado 80309, USA}
	\author{J. Gao}
	\affiliation{National Institute of Standards and Technology, Boulder, Colorado 80305, USA}
	\affiliation{Department of Physics, University of Colorado, Boulder, Colorado 80309, USA}
	\date{\today}
	
	\begin{abstract}
		We present a theoretical model and experimental characterization of a microwave kinetic inductance traveling-wave amplifier (KIT), whose noise performance, measured by a shot-noise tunnel junction (SNTJ), approaches the quantum limit. Biased with a dc current, the KIT operates in a three-wave mixing fashion, thereby reducing by several orders of magnitude the power of the microwave pump tone and associated parasitic heating compared to conventional four-wave mixing KIT devices. It consists of a 50 \ohm~artificial transmission line whose dispersion allows for a controlled amplification bandwidth. We measure $16.5^{+1}_{-1.3}$ dB of gain across a 2 GHz bandwidth with an input 1 dB compression power of -63 dBm, in qualitative agreement with theory. Using a theoretical framework that accounts for the SNTJ-generated noise entering both the signal and idler ports of the KIT, we measure the system-added noise of an amplification chain that integrates the KIT as the first amplifier. This system-added noise, $3.1\pm0.6$ quanta (equivalent to $0.66\pm0.15$ K) between 3.5 and 5.5 GHz, is the one that a device replacing the SNTJ in that chain would see. This KIT is therefore suitable to read large arrays of microwave kinetic inductance detectors and promising for multiplexed superconducting qubit readout.
		
	\end{abstract}
	
	\maketitle

	\section{INTRODUCTION}
	
	Is it possible to build a quantum limited microwave amplifier with enough gain, bandwidth and power handling to simultaneously read thousands of frequency-multiplexed superconducting resonators, like those in qubit systems or microwave kinetic inductance detectors (MKIDs)? When designed with resonant structures, Josephson junction-based parametric amplifiers have demonstrated high gain and quantum limited performances \cite{castellanos2008amplification,bergeal2010phase,Roch2012widely,mutus2013design,zhong2013squeezing,lecocq2017nonreciprocal,malnou2018optimal}. However, despite efforts to increase the bandwidth up to a few hundred megahertz via impedance engineering \cite{mutus2014strong,roy2015broadband}, or to increase the power handling up to a few hundred femtowatts via Kerr engineering \cite{frattini2017three,frattini2018optimizing,sivak2019kerr}, they still cannot read more than a handful of resonators simultaneously. When designed with nonresonant structures, i.e. transmission lines, Josephson traveling-wave parametric amplifiers (JTWPAs) have high gain over gigahertz bandwidth \cite{macklin2015near,white2015traveling,planat2020photonic}, but so far still exhibit similar power handling capabilities as their resonant counterparts. Recent studies \cite{Sivak2020Josephson,zorin2016Josephson,zorin2019flux} suggest that a three-wave mixing (3WM) JTWPA with a finely controlled and canceled Kerr nonlinearity should increase the device's power handling tenfold. Compelling experiments have yet to prove the feasibility of this approach, for which the JTWPA's design and fabrication increase in complexity. We propose to tackle this challenge using a different non-linear medium: starting with the intrinsic broadband and high power handling capabilities of a kinetic inductance traveling-wave amplifier (KIT) \cite{eom2012wideband}, we build a near-quantum-limited amplifier.
	
	The current limitations on microwave amplification affect many scientific endeavors. Although proof-of-principle ``quantum supremacy'' was demonstrated by a quantum computer containing a few tens of qubits \cite{arute2019quantum}, this number has to scale by at least an order of magnitude to run powerful quantum algorithms \cite{shor1994algorithms,grover1997quantum}. In the hunt for exoplanets, cameras with tens of thousands of MKID pixels are being built \cite{szypryt2017large}, and proposals to search for very light warm dark matter also necessitate the use of a great number of MKID pixels \cite{hochberg2016superconducting,hochberg2016detecting}. All these applications are either already limited by amplifier noise, or would greatly benefit from wideband, high gain, high power handling, quantum limited amplifiers.
	
	The KIT we present in this article is a step toward a practical, quantum-limited amplifier, whose bandwidth and power handling are compatible with high channel-count applications. Operating in a 3WM fashion, and fabricated out of a single layer of NbTiN, it consists of a weakly dispersive artificial transmission line \cite{chaudhuri2017broadband,zobrist2019wide}, for which we control the phase matched bandwidth with dispersion engineering. This limits spurious parametric conversion processes that otherwise degrade the power handling and noise performance. We measure an average gain of 16.5 dB over a 2 GHz bandwidth, and a typical 1 dB input compression power of -63 dBm within that bandwidth. Using a shot-noise tunnel junction (SNTJ) \cite{Spietz2003primary,spietz2006shot} we measure the added noise of a readout chain with the KIT as the first amplifier. To quote the \textit{true system-added} noise of the chain, i.e. the one that a device replacing the SNTJ in that chain would see, we develop a novel theoretical framework that accounts for the SNTJ-generated noise illuminating \textit{both} the signal and idler ports of the KIT. Failure to account for the idler port's noise makes the system-added noise look significantly better than its true value. We demonstrate a true system-added noise of $3.1\pm0.6$ quanta between 3.5 and 5.5 GHz, and estimate that the KIT alone is near-quantum-limited. It is the first time that the broadband noise properties of a KIT are fully characterized rigorously.
	
	\section{THEORY AND DESIGN}
	\label{sec:theo}
	
	KITs exploit the nonlinear kinetic inductance of a superconducting line to generate parametric interaction between pump, signal, and idler photons. In 3WM, a single pump photon converts into signal and idler photons, whereas four-wave mixing (4WM) converts two pump photons in this fashion. Operating a KIT with 3WM offers two key advantages over 4WM. First, as the pump frequency is far detuned from the amplification band, it is easily filtered, which is often necessary to avoid saturating the following amplifier. Second, it reduces the rf pump power because energy is extracted from dc power to convert pump photons, which avoids undesirable heating effects from the strong rf pump, including those happening in the packaging. More precisely, when biased with a dc current $I_d$, the KIT inductance per unit length L is \cite{vissers2016low}:
	\begin{equation} \label{eq:L}
	L = L_d(1+\epsilon I + \xi I^2 + \mathcal{O}(I^3)),
	\end{equation}
	where $I$ is the rf current, $L_d$ is the KIT inductance under dc bias, at zero rf current, $\epsilon=2I_d/(I_*^2+I_d^2)$ and $\xi=1/(I_*^2+I_d^2)$. The current $I_*$ sets the scale of the nonlinearity; it is typically $\sim10^3$ higher than that of Josephson devices, thereby conferring KITs $\sim10^4-10^6$ higher power handling capabilities than their Josephson equivalents. The term $\epsilon I$ permits 3WM, while $\xi I^2$ permits 4WM.
	
	Solving the coupled mode equations (CME) for a pump at frequency $\omega_p$, signal at $\omega_s$, and idler at $\omega_i$, such that $\omega_p =\omega_s+\omega_i$ yields the 3WM phase matching condition for exponential gain:
	\begin{equation} \label{eq:PM}
	\Delta_k = -\frac{\xi I_{p0}^2}{8}(k_p-2k_s-2k_i),
	\end{equation}
	see appendix \ref{app:CME}. Here, $\Delta_k=k_p-k_s-k_i$ with $k_j$, $j\in\{p,s,i\}$ the pump, signal and idler wavenumbers, and $I_{p0}$ is the rf pump amplitude at the KIT's input. In a non-dispersive transmission line $\Delta_k=0$, and thus equation \ref{eq:PM} can naturally be fulfilled over a very wide frequency range in KITs, where $I_{p0}\ll I_*$. Although desirable within the amplification bandwidth, it is undesirable outside of that bandwidth, where multiple parametric conversion processes take place \cite{vissers2016low,erickson2017theory}. These processes deplete the pump, thereby degrading the amplifier's power handling, and they also induce multiple conversion mechanisms at each frequency, thereby increasing the amplifier-added noise.
	\begin{figure}[!h]
		\centering
		\includegraphics[scale=0.45]{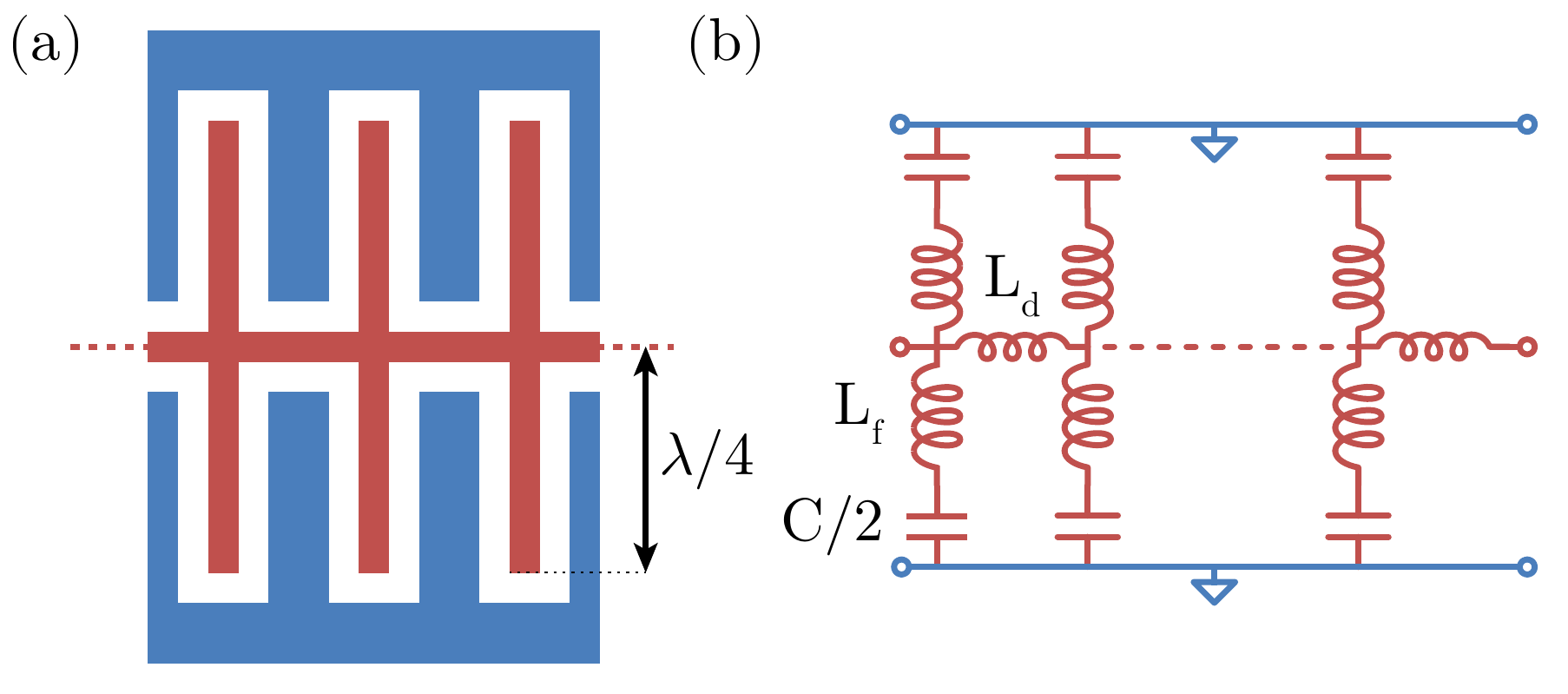} 
		\caption{Schematic of the KIT artificial transmission line. (a) Three cells in series (not to scale) are arranged in a CPW topology. The  highly inductive central line (red) is flanked by IDC fingers, which match it to 50 \ohm. Each finger constitutes a low-Q, quarter-wave resonator. (b) In the equivalent electrical circuit, each cell consists of a series inductance $L_d$ and two resonators with inductance $L_f$ and capacitance to ground $C/2$.} 
		\label{fig:schematicKIT}
	\end{figure}
	
\begin{figure}[h!]
		\centering
		\includegraphics[scale=0.45]{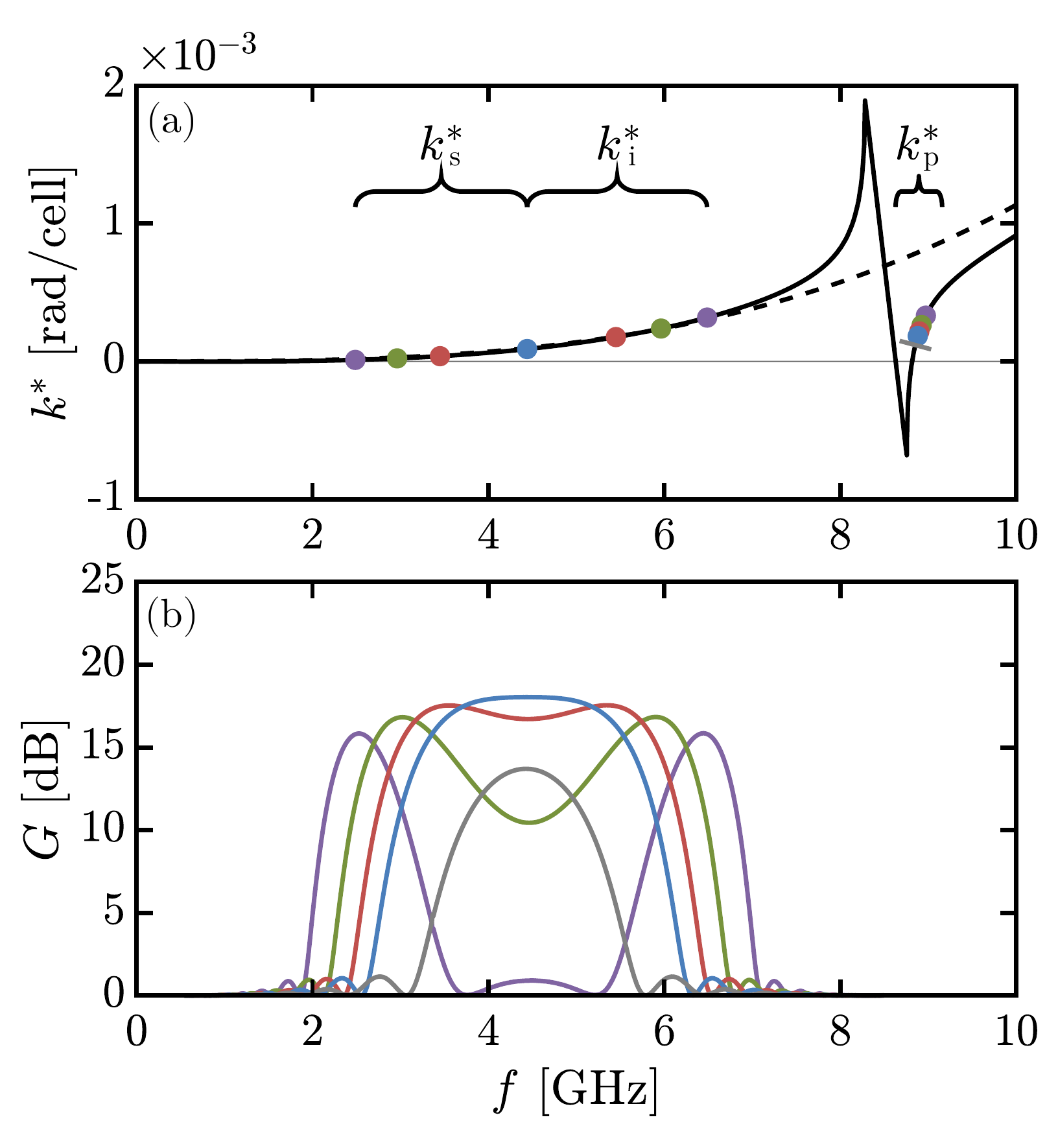} 
		\caption{Influence of the phase matching on the gain profile. (a) The nonlinear wavenumber $k^*$ is calculated as a function of frequency for the transmission line represented in Fig.~\ref{fig:schematicKIT} (dashed), and for a similar line, periodically loaded at $80$ \ohm~(black). The nonlinear part of four triplets $\{k_s,k_i,k_p\}$ that satisfy the phase matching condition (Eq.~\ref{eq:PM}) are indicated with colored dots: in purple $\omega_i-\omega_s=4$ GHz, green $\omega_i-\omega_s=3$ GHz, red $\omega_i-\omega_s=2$ GHz, and  blue $\omega_i=\omega_s$. Additionally, a gray line indicates a pump frequency for which phase matching is nowhere fulfilled (b) Solving the CME (Eqs.~\ref{eq:CME}), the signal power gain profile is calculated (Eq.~\ref{eq:gsdef}) for the related pump frequencies (the colors match with panel a). The KIT length is $N_c=6.6\times10^4$ cells.
		}
		\label{fig:PMgaintheo}
	\end{figure}

	While in conventional traveling-wave amplifiers dispersion engineering prevents only pump harmonic generation, we suppress all unwanted parametric conversion processes by designing our KIT as a weakly dispersive artificial transmission line. Originally developed to have the KIT matched to 50 \ohm\space\cite{chaudhuri2017broadband}, this line consists of a series of coplanar waveguide (CPW) sections, or \textit{cells}, each with inductance $L_d$, flanked by two interdigitated capacitor (IDC) fingers that form the capacitance to ground $C$, such that $Z=\sqrt{L_d/C}=50$ \ohm, see Fig.~\ref{fig:schematicKIT}a. Each IDC finger then constitutes a low-Q quarter-wave resonator, with capacitance $C_f=C/2$ and inductance $L_f$ (see Fig.~\ref{fig:schematicKIT}b), set by the finger's length. In practice, we choose $\omega_f=1/\sqrt{L_f C_f} = 2\pi\times36$ GHz, and $Q=1/Z\sqrt{L_f/C_f}=3.3$, to generate a slight dispersion at low frequencies, where the pump, signal and idler modes lie. The dashed line in Fig.~\ref{fig:PMgaintheo}a shows the dispersive part of the wavenumber $k^* = k - k_0$ with $k_0=\omega\sqrt{L_dC}$, as a function of frequency. It is calculated by cascading the ABCD matrices of the KIT cells, see appendix \ref{app:ABCD}. As it slightly deviates from zero, no triplet $\{k_s,k_i,k_p\}$ can satisfy Eq.~\ref{eq:PM}.
	
	To retrieve phase-matching over a desired bandwidth, we engineer another dispersion feature by increasing the line impedance periodically on short sections. This feature creates a resonance in the line's phase response (and a stopband in the line's amplitude response), at a frequency $\omega_l$ controlled by the loading periodicity \cite{chaudhuri2017broadband,pozar2011microwave}. Figure \ref{fig:PMgaintheo}a  shows $k^*$ in a line periodically loaded at $80$ \ohm, with $\omega_l = 2\pi\times 8.5$ GHz. Because the nonlinear wavenumber close to resonance varies sharply, there exists values of $k^*_p$ (above $\omega_l$) for which we can find triplets $\{k_s,k_i,k_p\}$ that satisfy Eq.~\ref{eq:PM} (examples of their nonlinear parts are shown in colored dots). A slight variation of the pump frequency $\omega_p$ significantly affects which pair of signal and idler frequencies is phase matched.
	
	At these matched frequencies, the 3WM gain grows exponentially with $k_p$ (in radians per cell), with the KIT length, and with $\delta_L$, the relative inductance modulation amplitude generated by the rf current and scaling with $I_d$. More precisely, the phase matched, small signal power gain can be written as:
	\begin{equation}
	G_s \simeq \cosh^2\left(\frac{1}{8}\delta_L k_pN_c\right),
	\label{eq:Gs}
	\end{equation}
	where $N_c$ is the total number of cells, see appendix \ref{app:3WMG}. Typically, when operating our KIT, $\delta_L\sim7\times10^{-3}$; thus, with $L_d\sim50$ pH/cell (see Sec.~\ref{sec:exp}), we need $N_c>5\times10^4$ to get $G_s>15$ dB at $\omega_p\sim2\pi\times9$ GHz.
	
	Since maximum gain is achieved with phase-matching, $\omega_p$ influences the gain profile. To calculate this profile, we insert the dispersion relation $k(\omega)$ into the CME, and solve them numerically, see appendix \ref{app:ABCD}. Figure \ref{fig:PMgaintheo}b shows signal power gain profiles, calculated for the pump frequencies represented in Fig.~\ref{fig:PMgaintheo}a. As expected, the gain is maximal at the signal and idler frequencies for which the triplets $\{k_s,k_i,k_p\}$ satisfy Eq.~\ref{eq:PM}. When these frequencies are far apart, there is a region in between where phase matching is not sufficient and the gain drops. By reducing $\omega_p$, we can lower the distance between phase-matched signal and idler frequencies, and therefore obtain a wideband, flat gain region. Further reducing $\omega_p$, we reach the value where phase-matched signal and idler frequencies are equal, beyond what phase matching is nowhere fulfilled anymore, and the gain drops. Fundamentally, the wideband nature of the gain depends on the convexity of the dispersion relation, and therefore on the fingers' length and capacitance to ground. As $\omega_f$ or Q increase, $k^*$ is less convex, and thus closer to a broadband, phase-matched situation, but at the cost of allowing extra parametric processes to arise.
	
	\section{EXPERIMENTAL REALIZATION}
	\label{sec:exp}
	
\begin{figure}[!h]
		\centering
		\includegraphics[scale=0.7]{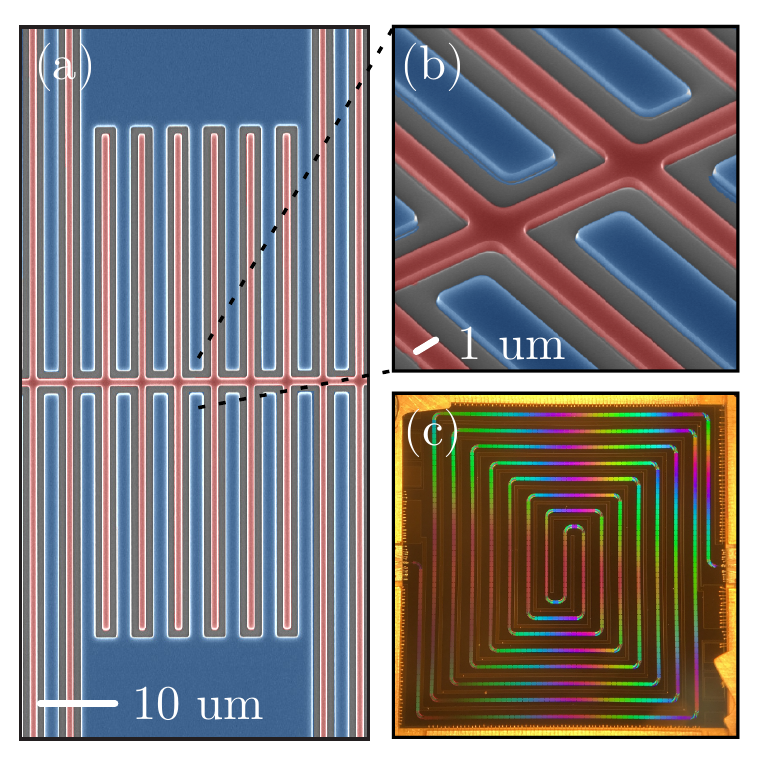} 
		\caption{Micrographs of the KIT. (a) The transmission line (false color red) is periodically loaded with shorter IDC fingers. (b) Line and fingers are 1 \micron wide, and each cell is 5 \micron long. (c) The overall KIT is laid out in a spiral configuration on a $2\times2$ cm chip, and clamped on a copper packaging.} 
		\label{fig:KITfab}
	\end{figure}

	Because the kinetic inductance nonlinearity is weak, in practice a KIT comprises a transmission line tens of centimeters long. To maximize this nonlinearity, and to minimize its length, the line as well as IDC fingers are made 1 \micron wide, and each unit cell is 5 \micron long, see Fig.~\ref{fig:KITfab}a and b. Fabricated from a 20 nm thick NbTiN layer on 380 \micron high-resistivity intrinsic silicon via optical lithography, it yields $I_*\sim7$ mA, and a sheet inductance $\sim10$ $\mathrm{pH}/$square. Thus, $L_d\sim50$ pH, and in order to retrieve $Z=50$ \ohm, each finger is made 102 \micron long. The loading (Fig.\ref{fig:KITfab}a) consists of cells with shorter fingers (33.5 $\upmu\mathrm{m}$, $Z=80$ \ohm), arranged periodically to generate a resonance at 8.5 GHz, thereby positioning the pump frequency in a way compatible with our filtering capabilities.
	
	We lay out the line in a spiral topology, on a $2\times2$ cm chip (Fig.~\ref{fig:KITfab}c), which contains $N_c=6.6\times10^4$ cells, equivalent to 33 cm. To avoid spurious chip and microstrip modes, electrical grounding inside the chip is ensured with spring-loaded pogo pins, descending from the top copper lid of the packaging, and contacting the chip between the line traces, see appendix \ref{app:packaging}. The pogo pins also improve the chip-to-packaging thermal link, which otherwise mostly relies on wire-bonds.
	
\begin{figure}[!h]
		\centering
		\includegraphics[scale=0.33]{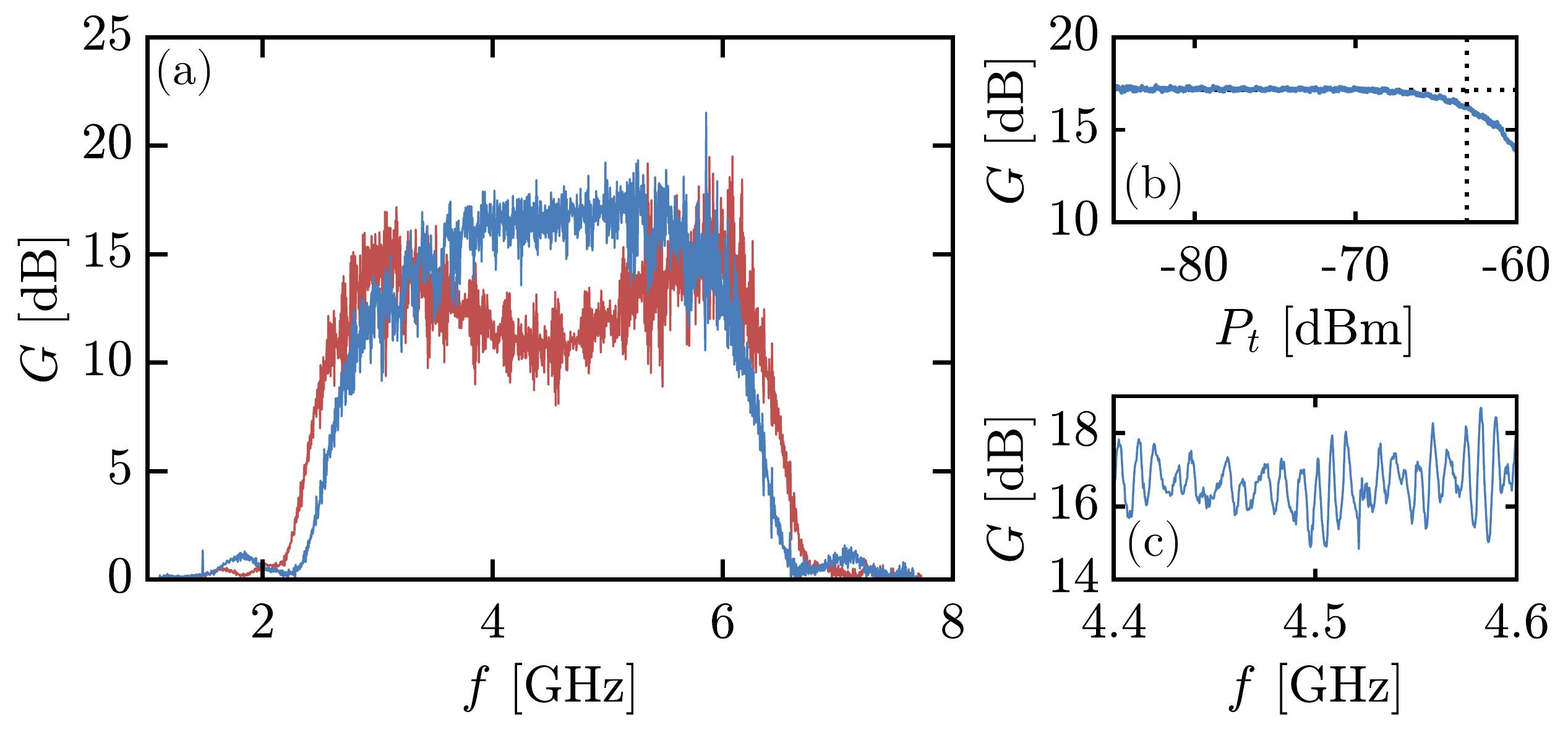} 
		\caption{Amplification properties of the KIT. (a) The power gain is measured with a vector network analyzer (VNA), when $\omega_p=2\pi\times8.895$ GHz (blue), and $\omega_p=2\pi\times8.931$ GHz (red). (b) The gain of a probe tone at 4.5 GHz compresses as the tone's power increases. At $P_\mathrm{-1dB}=-63$ dBm, referred to the KIT's input, the gain has lowered by 1 dB from its small signal value. It is measured for $\omega_p=2\pi\times8.895$ GHz. (c) At the same pump frequency, a close-up on the small signal gain around 4.5 GHz shows ripples with 8 MHz characteristic frequency.}
		\label{fig:GBWPHR}
	\end{figure}

	In a first experiment, we measure the gain, bandwidth, and power handling of the KIT when cooled to $\sim30$ mK. The KIT is mounted as the sole amplifier in the chain, thereby facilitating comparison of its gain profile to the theoretical profiles, and revealing the gain ripples of the isolated KIT, which otherwise also depend on the return loss of external components. Two mandatory bias tees at the KIT input and output ports combine dc and rf currents. Figure \ref{fig:GBWPHR}a shows KIT gain profiles, acquired at two different pump frequencies. The current amplitudes are $I_d=1.5$ mA, and $I_{p0}=160$ \microamp ~($-29$ dBm in power, calibrated \textit{in situ} by comparing dc and rf nonlinear phase shifts, see appendix \ref{app:ps}). For the higher pump frequency, the gain drops in the middle of the amplification bandwidth. For the lower one, the gain profile is flatter, with an average value of $16.5^{+1}_{-1.3}$ dB between 3.5 and 5.5 GHz, where the subscript and superscript denote the amplitude of the gain ripples. Both profiles agree qualitatively with behaviors explained in Sec.~\ref{sec:theo}. The gain ripples have a $8$ MHz characteristic frequency (see Fig.~\ref{fig:GBWPHR}c), equivalent to $62.5$ cm in wavelength (the phase velocity being $v_p=1/\sqrt{L_dC}\sim1000$ cell per nanosecond), or about twice the KIT length. We thus attribute their presence to a mismatch in impedance between KIT and microwave feed structures before and after the KIT.  This mismatch results in a pump standing wave pattern, which influences the signal amplification, depending on its frequency. Figure \ref{fig:GBWPHR}b shows the gain at 4.5 GHz (obtained at the lower pump power), as a function of $P_t$, the input power of a probe tone. The gain compresses by 1 dB from its small signal value for $P_\mathrm{-1dB} = -63$ dBm, about $7$ dB lower than theoretical predictions, see appendix \ref{app:ph}. This discrepancy, suggesting substantial room for improvement, may be due to effects not included in our model, such as standing wave patterns, or defects in the line, which locally lower $I_*$. Nonetheless, $P_\mathrm{-1dB}$ is already about $30$ dB higher than JTWPAs \cite{macklin2015near,white2015traveling,planat2020photonic}, and about $10$ dB less than 4WM KITs with similar gain (see appendix \ref{app:kitcomparisons}). This KIT is therefore suitable to read thousands of frequency multiplexed MKIDs, that use drive tone powers typically around $-90$ dBm \cite{zobrist2019wide,vissers2016low} or even more qubits, whose readout involves powers substantially less than $-90$ dBm.
	
	As in any phase-insensitive traveling-wave and resonant parametric amplifier, the practical, usable bandwidth, is half of the presented amplification bandwidth. It is the bandwidth in which signals coming from microwave devices can be phase-insensitively amplified. The other half, barring the idler frequencies, contains a copy of signals in the first half. That is why the gain in Fig.~\ref{fig:GBWPHR}a is nearly symmetric about the half pump frequency ($\sim4.5$ GHz). The asymmetry - gain and ripples marginally bigger above the half pump frequency - originates from a frequency dependent saturation power. In fact, higher frequencies possess a higher saturation power, see appendix \ref{app:gainasym}. We see this effect here because we chose a signal power close to $P_\mathrm{-1dB}$ in order to maximize the signal-to-noise ratio in this measurement, where the KIT remains the sole amplifier.
	
	At higher dc current bias (bounded by the dc critical current of the transmission line, $\sim2.4$ mA in our device), lower rf pump power can be used to obtain equivalent small signal gain, at the cost of a reduced 1 dB compression power. Conversely, reducing $I_d$ and increasing $I_{p0}$ improves power handling capabilities, but the gain is then limited by a superconductivity breaking phenomenon. We suspect the presence of weak links \cite{bockstiegel2014development}, and we are currently investigating the line breakdown mechanism.

	\section{NOISE PERFORMANCE}
	\label{sec:noise}
	
	The combined gain, bandwidth, and power handling performance are promising, provided that the KIT also presents competitive noise performance. Measuring this noise is a topic of current interest \cite{eom2012wideband,ranzani2018kinetic,zobrist2019wide}, and we execute the task using a self-calibrated shot-noise tunnel junction (SNTJ) \cite{Spietz2003primary,spietz2006shot}. We measure the output spectral density, whose power depends on the chain's gain and loss. The SNTJ acts as a dynamic variable noise source, allowing for a continuous sweep in input noise temperature by sweeping its bias voltage, and our measurement scans the entire KIT bandwidth.
	
	\begin{figure}[!h]
		\centering
		\includegraphics[scale=0.52]{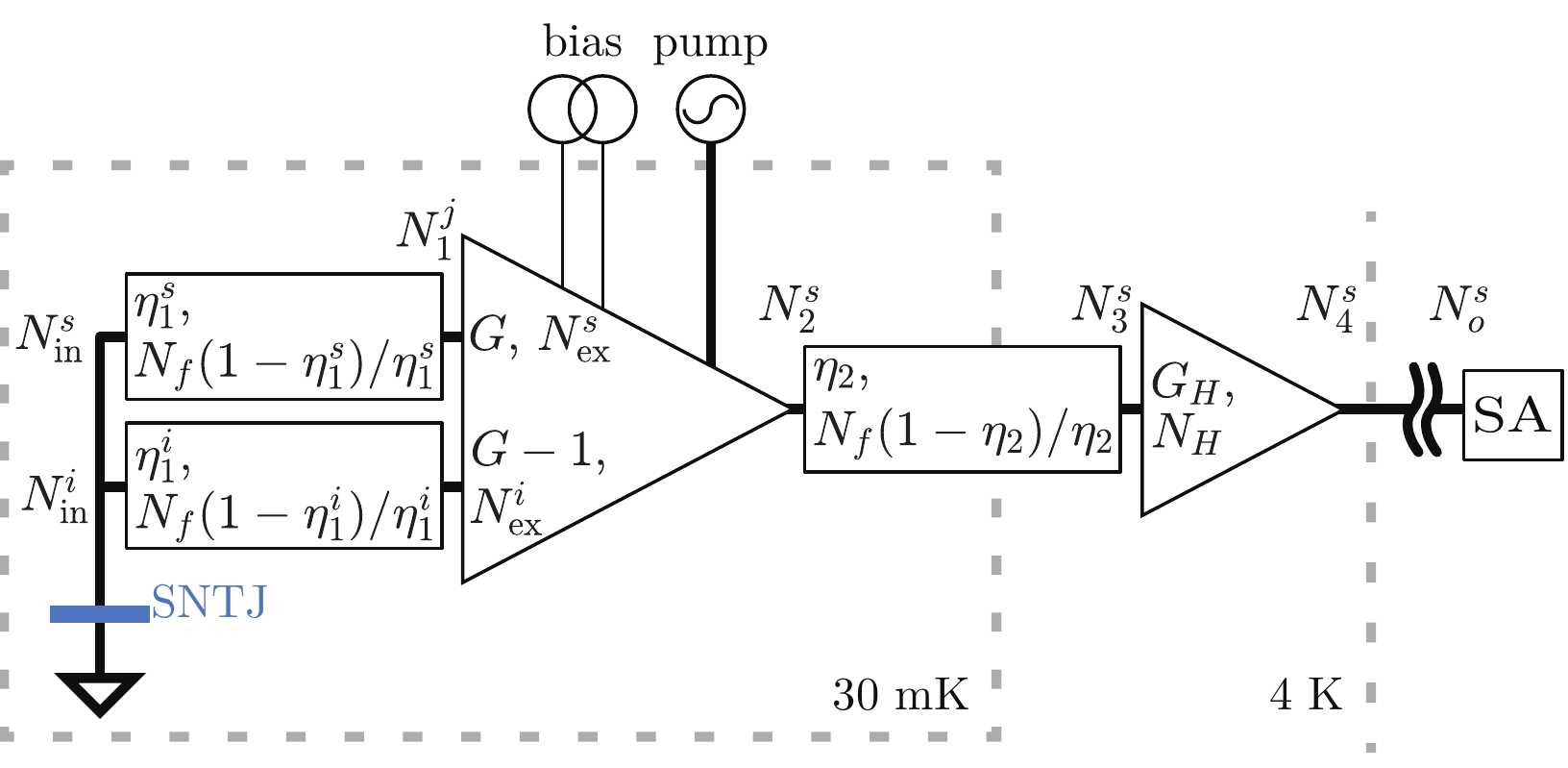} 
		\caption{Schematic of the noise measurement setup. Each component is labeled with its gain or transmission efficiency, and with its input-referred added noise. Calibrated noise $N_\mathrm{in}^s$ ($N_\mathrm{in}^i$) is generated by the SNTJ at the signal (idler) frequency. It is routed to the KIT with transmission efficiency $\eta_1^s$ ($\eta_1^i$), i.e. it undergoes a beamsplitter interaction and part of it is replaced with vacuum noise whose value $N_f=0.5$ quanta. At the KIT's input, the noise is $N_1^j$, $j\in\{s,i\}$. On the signal-to-signal path, the KIT then adds $N_\mathrm{ex}^s$ quanta of excess noise and has a gain $G$; on the idler-to-signal path, the KIT adds $N_\mathrm{ex}^i$ quanta of excess noise and has a gain $G-1$. Noise $N_2^s$ at the KIT's output is then routed with efficiency $\eta_2$ to the HEMT (input noise $N_3^s$). With gain $G_H$ and added noise $N_H$, it further directs the noise $N_4^s$ to room temperature components. Amplification and loss at room temperature are excluded from our schematic but not our analysis. The noise reaching the spectrum analyzer (SA) is $N_o^s$. The full setup is described in appendix \ref{app:setupnoise}.} 
		\label{fig:setupnoise}
	\end{figure}
	\begin{figure*}[t!] 
		\centering
		\includegraphics[scale=0.6]{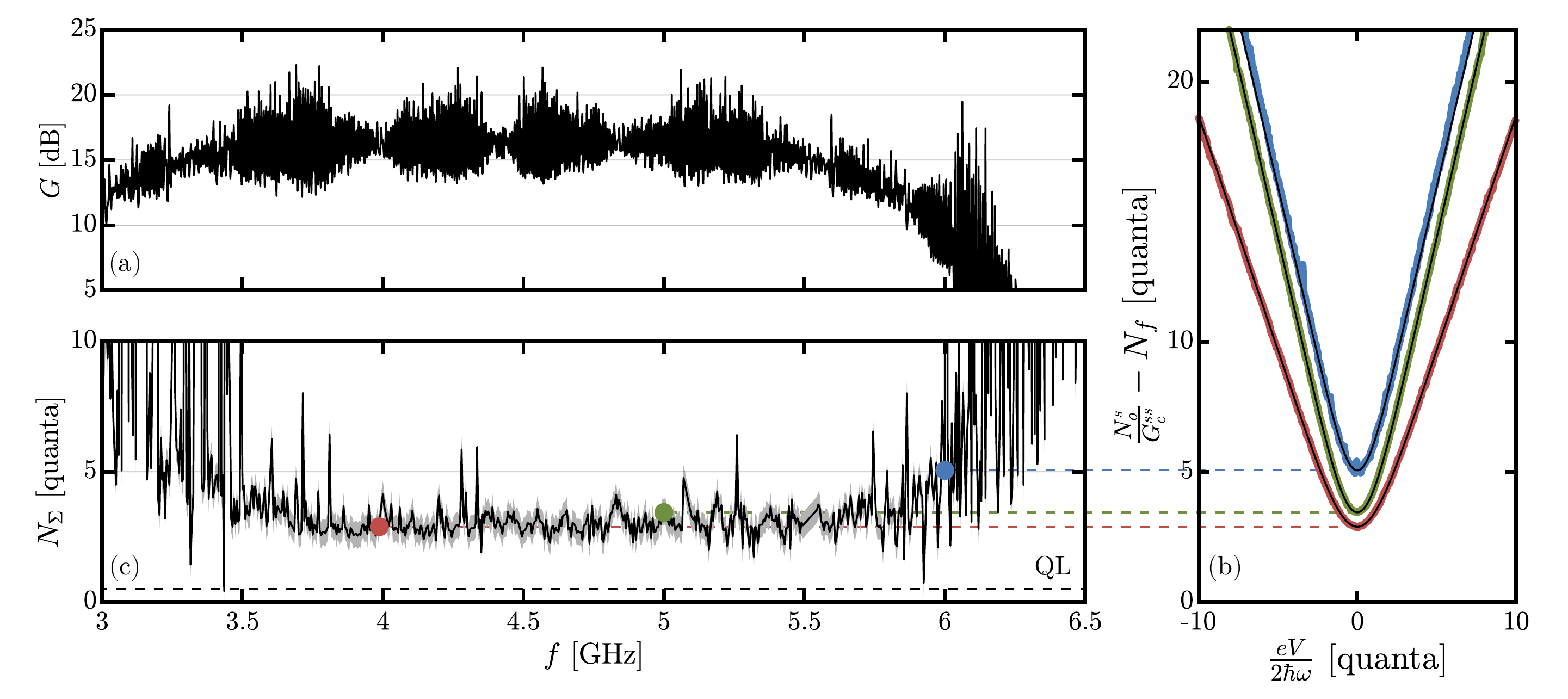}
		\caption{System-added noise measurement of a microwave amplification chain with a KIT as first amplifier. (a) The gain's frequency dependence is measured with a VNA (with a 1 kHz intermediate frequency bandwidth). (b) The output noise $N_o^s$ is measured with a SA (5 MHz resolution bandwidth, comparable to typical resonant amplifier's bandwidth), while varying the SNTJ dc voltage bias V. Fitting the whole output noise response, we obtain the frequency dependent system-added noise $N_\Sigma$ and the chain's signal-to-signal gain $G_c^{ss}$. We divide $N_o^s$ by $G_c^{ss}$ to refer it to the KIT input, and subtract the zero bias noise value $N_f=0.5$, so that $N_\Sigma$ (panel c) visually matches the zero bias value of $N_o^s$ (see appendix \ref{app:sn}). Three colored curves indicate output noises at 4, 5, and 6 GHz, with fits superimposed in black lines. (c) Data from the output noise spectra are compiled to form $N_\Sigma$. Uncertainties are indicated by the gray area surrounding the black line. They predominantly come from the fit of the curves in (b). The quantum limit (QL) on amplifier-added noise is indicated by the dashed black line.} 
		\label{fig:KITnoise}
	\end{figure*}

	
	Because the SNTJ is a wideband noise source, it illuminates the KIT at \textit{both} its signal and idler frequency inputs; these input noises are then parametrically amplified. We thus model the KIT as a three-port network: two inputs at $\omega_s$ and $\omega_i$, and one output at $\omega_s$. Figure \ref{fig:setupnoise} schematizes the entire amplification chain, where we have labeled the gain or transmission efficiency, and input-referred added noise of each component. The KIT power gain on the signal-to-signal path is $G$, while its power gain on the idler-to-signal path is $G-1$ \cite{caves1982quantum}. The output noise at the signal frequency (in units of photons) measured on the spectrum analyzer (SA) can then be written as
	\begin{equation} 
	\label{eq:No}
	N_o^s = G_c^{ss}(N_\mathrm{in}^s + N_\mathrm{eff}^s) + G_c^{si}(N_\mathrm{in}^i + N_\mathrm{eff}^i), 
	\end{equation}
	where $N_\mathrm{in}^s$ ($N_\mathrm{in}^i$) is the SNTJ-generated noise at the signal (idler) frequency, $G_c^{ss}$ ($G_c^{si}$) is the signal-to-signal (idler-to-signal) gain of the entire chain from SNTJ to SA, and $N_\mathrm{eff}^s$ ($N_\mathrm{eff}^i$) is the effective signal-to-signal (idler-to-signal) path KIT-excess noise, see appendix \ref{app:noise}. When varying $N_\mathrm{in}^s$ and $N_\mathrm{in}^i$ we retrieve $N_\mathrm{eff}^s$ and $N_\mathrm{eff}^i$, equal to zero for a quantum-limited amplifier.
	
	The system-added noise that a device replacing the SNTJ would see is therefore
	\begin{equation} \label{eq:Nsigma}
	N_\Sigma = N_\mathrm{eff}^s + \frac{G_c^{si}}{G_c^{ss}}(N_f+N_\mathrm{eff}^i),
	\end{equation}
	where $N_f=0.5$ quanta is the vacuum noise (provided that the idler port of this device is cold); for a high gain, phase-insensitive quantum-limited amplifier $N_f$ is the minimum added noise \cite{caves1982quantum}. Failure to account for the additional change in idler noise at the amplifier's input leads to an underestimate of the system-added noise by about a factor two, thereby making it look significantly better than its true value, given by Eq.~\ref{eq:Nsigma}, see appendix \ref{app:noisecorrection}.
	
	In practice, we measure $N_o^s$ while varying simultaneously $N_\mathrm{in}^s$ and $N_\mathrm{in}^i$ using the SNTJ voltage bias (Fig.~\ref{fig:KITnoise}b). We then fit to obtain $G_c^{ss}$, $G_c^{si}$, $N_\mathrm{eff}^s$ and $N_\mathrm{eff}^i$ (see appendix \ref{app:sn}), and form $N_\Sigma$ with Eq.~\ref{eq:Nsigma}. Figure \ref{fig:KITnoise}c presents the system-added noise $N_\Sigma$, measured over the KIT's amplification bandwidth. In this experimental configuration, the KIT gain is $G=16.6^{+1.8}_{-3.1}$ dB between 3.5 and 5.5 GHz (Fig.~\ref{fig:KITnoise}a), stable over the acquisition time ($\sim12$ hrs). In that bandwidth, $N_\Sigma=3.1\pm0.6$ quanta. It is an unprecedented broadband, true system-added noise performance (see appendix \ref{app:kitcomparisons}).
	
	This performance depends on the intrinsic signal (idler) KIT-excess noise $N_\mathrm{ex}^s$ ($N_\mathrm{ex}^i$), but also on the chain's transmission efficiencies as well as on the HEMT-added noise $N_H$. More precisely, when $\{G,N_H\}\gg1$, Eq.~\ref{eq:Nsigma} becomes:
	\begin{equation}
	N_\Sigma = \frac{N_\mathrm{ex}^s+N_\mathrm{ex}^i}{\eta_1^s} + \frac{2(1-\eta_1^s)N_f}{\eta_1^s} + \frac{N_H}{\eta_2 G \eta_1^s} + N_f.
	\label{eq:Nsigma_simple}
	\end{equation}
	From left to right, the first three terms in the right-hand side represent the contribution to $N_\Sigma$ from the KIT alone, from the lossy link between the SNTJ and the KIT, and from the HEMT-added noise; the last term is the minimum half quantum of added noise that a quantum-limited amplifier must add \cite{caves1982quantum}. Measuring the individual loss of the chain's components, we estimate the value of the transmission efficiencies to be $\eta_1^s=0.57\pm0.02$ and $\eta_2=0.64\pm0.10$; in addition, by measuring the system-added noise of the amplification chain with the KIT turned off, we estimate $N_H=8\pm1$ quanta, see appendix \ref{app:lossbudget}. With these additional information, we \textit{estimate} that the overall KIT-excess noise is $N_\mathrm{ex}^s + N_\mathrm{ex}^i=0.77\pm0.40$ quanta, suggesting that the KIT alone operates near the quantum limit.
	
	Several strategies can improve the system-added noise. First, increasing the transmission efficiencies would have a major impact, because all the system-dependent noise contributions are enlarged by $1/\eta_1^s$. For example, $\eta_1^s=0.8$ would yield $N_\Sigma=1.6$ quanta. To that end, we are currently developing lossless superconducting circuits (bias tees, directional couplers and low-pass filters) that can be integrated on-chip with the KIT. Second, increasing the KIT gain $G$ would reduce the HEMT contribution to $N_\Sigma$ (third term in Eq.~\ref{eq:Nsigma_simple}). Here, this contribution is estimated at $0.5\pm0.3$ quanta. The solutions to achieve higher gain directly follow from Eq.~\ref{eq:Gs}: higher pump power (i.e. increase $\delta_L$), longer line (increase $N_c$), or higher inductance per unit cell (increase $k_p$). All represent non-trivial challenges, starting with better understanding of the line breakdown mechanism \cite{bockstiegel2014development}. If it comes from imperfections in the line (weak links), a higher resolution fabrication process, like electron beam lithography, may improve the performance of the device. Also, running the amplifier at higher gain will require better damping of the gain ripples, whose amplitude grows with gain. Finally, we are investigating the origin of the remaining excess noise $N_\mathrm{ex}^s+N_\mathrm{ex}^i$. It may be due to parasitic chip heating or two-level system noise \cite{gao2008experimental}.

	\section{CONCLUSION}
	
	It is possible to build a microwave amplifier with broadband and near-quantum-limited performance without sacrificing power handling capability. Engineering the phase-matched bandwidth is key because it suppresses spurious parametric conversion processes. We demonstrate this idea on a KIT, whose combined gain, bandwidth, power handling, and noise performance are fully characterized. In addition, we develop a theoretical framework adapted to noise measurements performed with wideband noise sources. Using a SNTJ we evaluate the true system-added noise of an amplification containing the KIT as the first amplifier. The KIT itself is estimated to be near-quantum-limited, therefore it has the potential to initiate a qualitative shift in the way arrays of superconducting detectors, such as MKIDs, process information, moving them into a quantum-limited readout regime.
	
	\section*{Acknowledgments} 
	
	We thank Kim Hagen for his help in the design and fabrication of the KIT packaging, and Felix Vietmeyer and Terry Brown for their help in the design and fabrication of room temperature electronics. We acknowledge useful discussions with John Teufel, Gangqiang Liu and Vidul Joshi. Certain commercial materials and equipment are identified in this paper to foster understanding. Such identification does not imply recommendation or endorsement by the National Institute of Standards and Technology, nor does it imply that the materials or equipment identified are necessarily the best available for the purpose. This work was supported by the NIST Innovations in Measurement Science Program, as well as NASA, under grant NNH18ZDA001N-APRA.
	
	\appendix
	
	\section{COUPLED-MODE THEORY OF A DC-BIASED KIT}
	
	\subsection{Coupled-mode equations}
	\label{app:CME}
	
	The phase matching condition for exponential gain, Eq.~\ref{eq:PM} is obtained by solving the CME while pumping in a 3WM fashion, i.e. such that $\omega_p =\omega_s+\omega_i$, and in the presence of 3WM and 4WM terms see, Eq.~\ref{eq:L}. The CME relate the current amplitudes $I_j$, $j\in\{p,s,i\}$ at the frequencies $\omega_j$, $j\in\{p,s,i\}$; they are obtained by injecting equation \ref{eq:L} into the telegrapher's equations, and by operating the harmonic balance (HB) with only these three frequencies.
	
	More precisely, the telegrapher's equations in a lossless transmission line relate current I and voltage V as:
	\begin{equation}
	\begin{aligned}
	-\frac{\partial I}{\partial x} &= C \frac{\partial V}{\partial t}\\
	-\frac{\partial V}{\partial x} &= L \frac{\partial I}{\partial t},
	\end{aligned}
	\label{eq:teleg}
	\end{equation}
	with x a length per unit cell. Injecting Eq.~\ref{eq:L} into Eqs.~\ref{eq:teleg}, we obtain:
	\begin{equation}
	v_p^2\frac{\partial^2I}{\partial x^2}-\frac{\partial^2I}{\partial t^2} = \frac{\partial}{\partial t^2}\left(\frac{1}{2}\epsilon I^2 + \frac{1}{3} \xi I^3\right),
	\label{eq:prop}
	\end{equation}
	with $v_p=1/\sqrt{C L_d}$ the phase velocity. To solve Eq.~\ref{eq:prop} we perform the HB: we \textit{assume} that the current in the transmission line is a sum of three terms at three different frequencies:
	\begin{equation}
	\begin{aligned}
	I = \frac{1}{2}\big(&I_p(x)e^{i(k_px-\omega_pt)}\\
	+&I_s(x)e^{i(k_sx-\omega_st)}\\
	+&I_i(x)e^{i(k_ix-\omega_it)} + c.c\big),
	\end{aligned}
	\label{eq:I}
	\end{equation}
	and we then keep only the mixing terms from Eq.~\ref{eq:prop} that emerge at these frequencies. This approach is valid in our case, because the phase matching bandwidth is limited by dispersion engineering (see appendix \ref{app:ABCD}), and thus mostly these three frequencies are able to mix together. Under the slow-varying envelope approximation, $\lvert d^2 I_j/dx^2 \rvert \ll \lvert k_j d I_j/dx \rvert$ for $j\in\{p,s,i\}$, the left hand side of Eq.~\ref{eq:prop} yields:
	\begin{equation}
	v_p^2\frac{\partial^2I}{\partial x^2}-\frac{\partial^2I}{\partial t^2} = v_p^2\sum_{j = p,s,i}ik_j\frac{d I_j}{dx}e^{i(k_jx-\omega_jt)}+c.c.
	\end{equation}
	Using $\omega_p =\omega_s+\omega_i$, we collect terms at $\omega_j$, $j\in\{p,s,i\}$ in the right hand side (rhs) and form the CME:
	\begin{equation}
	\begin{aligned}
	\frac{dI_p}{dx} &= \frac{i k_p \epsilon}{4} I_s I_i e^{-i \Delta_k x} + \frac{i k_p \xi}{8}I_p(\lvert I_p \rvert^2 + 2\lvert I_s \rvert^2 + 2\lvert I_i \rvert^2)\\
	\frac{dI_s}{dx} &= \frac{i k_s \epsilon}{4} I_p I_i^* e^{i \Delta_k x} + \frac{i k_s \xi}{8}I_s(2\lvert I_p \rvert^2 + \lvert I_s \rvert^2 + 2\lvert I_i \rvert^2)\\
	\frac{dI_i}{dx} &= \frac{i k_i \epsilon}{4} I_p I_s^* e^{i \Delta_k x} + \frac{i k_i \xi}{8}I_i(2\lvert I_p \rvert^2 + 2\lvert I_s \rvert^2 + \lvert I_i \rvert^2),
	\end{aligned}
	\label{eq:CME}
	\end{equation}
	with $\Delta_k=k_p-k_s-k_i$. The phase matching condition, Eq.~\ref{eq:PM}, is found for a strong pump, where $\{\lvert I_s\rvert, \lvert I_i \rvert\}\ll\lvert I_p\rvert$. Assuming the pump undepleted, $\lvert I_p(x)\rvert=I_{p0}$, Eqs.~\ref{eq:CME} rewrite:
	\begin{equation}
	\begin{aligned}
	\frac{dI_p}{dx} &= \frac{i k_p \xi}{8}I_p I_{p0}^2\\
	\frac{dI_s}{dx} &= \frac{i k_s \epsilon}{4} I_p I_i^* e^{i \Delta_k x} + \frac{i k_s \xi}{4}I_s I_{p0}^2\\
	\frac{dI_i}{dx} &= \frac{i k_i \epsilon}{4} I_p I_s^* e^{i \Delta_k x} + \frac{i k_i \xi}{4}I_i I_{p0}^2,
	\end{aligned}
	\label{eq:CMEstrong}
	\end{equation}
	which results in $I_p(x)=I_{p0}\exp{(i\xi k_p I_{p0}^2 x /8)}$. Signal and idler amplitudes are then searched with the form: $I_j(x)= \Tilde{I}_{j}(x) \exp{(i\xi I_{p0}^2k_j x/4)}$, $j\in\{s,i\}$. Equations \ref{eq:CMEstrong} then yield:
	\begin{equation}
	\begin{aligned}
	\frac{d \Tilde{I}_s}{dx} &= \frac{i k_s \epsilon}{4} I_{p0} \Tilde{I}_i^* e^{i \Delta_\beta x}\\
	\frac{d \Tilde{I}_i}{dx} &= \frac{i k_i \epsilon}{4} I_{p0} \Tilde{I}_s^* e^{i \Delta_\beta x},
	\end{aligned}
	\label{eq:CMEsi}
	\end{equation}
	with $\Delta_\beta=\Delta_k+\frac{\xi I_{p0}^2}{8}(k_p-2k_s-2k_i)$ and $\Delta_k=k_p-k_s-k_i$. The system of equations \ref{eq:CMEsi} has known solutions \cite{boyd2019nonlinear}. In particular, when phase matching is achieved, i.e. $\Delta_\beta=0$, we obtain:
	\begin{equation}
	\begin{aligned}
	\Tilde{I}_s &= \cosh{(g_3x) \Tilde{I}_{s0}}\\
	\Tilde{I}_i &= i\sqrt{\frac{k_i}{k_s}}\sinh{(g_3x) \Tilde{I}_{s0}},
	\end{aligned}
	\label{eq:siPM}
	\end{equation}
	with $g_3=\frac{\epsilon I_{p0}}{4}\sqrt{k_i k_s}$, and with initial conditions $I_s(0)=I_{s0}$ and $I_i(0)=0$. The signal power gain
	\begin{equation}
	G_s(x) = \left\lvert\frac{I_s(x)}{I_{s0}}\right\rvert^2
	\label{eq:gsdef}
	\end{equation}
	is then exponential with $x$: $G_s=\cosh^2(g_3x)$.
	
	\subsection{3WM gain}
	\label{app:3WMG}
	
	We can re-write $g_3$ as a function of more meaningful quantities. In fact, the linear inductance $L$ exposed in Eq.~\ref{eq:L} also writes:
	\begin{equation}
	\begin{aligned}
	L &= L_0\left(1+\frac{I_d^2}{I_*^2}\right)\left(1+\frac{2I_dI}{I_*^2+I_d^2} + \frac{I^2}{I_*^2+I_d^2}\right)\\
	&\simeq L_d\left(1+\frac{2I_dI}{I_*^2+I_d^2}\right),
	\end{aligned}
	\end{equation}
	when $I\ll I_d$, and with $L_d = L_0(1+I_d^2/I_*^2)$. Here, $L_0$ is the bare linear inductance. It is the one directly derived from the sheet kinetic inductance and the geometry of the line, while $L_d$ is the inductance per unit length under dc bias. Because $I_d^2\ll I_*^2$, for design purposes $L_d\sim L_0$. In the strong pump regime $I=I_{p0}$, therefore, $2I_dI/(I_*^2+I_d^2) = \epsilon I_{p0}$; assuming $k_s=k_i\simeq k_p/2$, we therefore get:
	\begin{equation}
	G_s(x) \simeq \cosh^2\left(\frac{1}{8}\delta_Lk_p x\right),
	\end{equation}
	where 
	\begin{equation}
	\delta_L = \frac{L-L_d}{L_d} = \frac{2 I_dI_{p0}}{I_*^2 + I_d^2}
	\end{equation}
	is the relative inductance variation due to $I_{p0}$. With $I_*=7$ mA, and typical values: $I_d=1.5$ mA (limited by the dc critical current, $\sim2.4$ mA, value specific to our NbTiN film and to the line's width) and $I_{p0}=I_*/60$, we get $\delta_L=6.8\times10^{-3}$.
	
	\subsection{Pump phase shift}
	\label{app:ps}
	
	From the phase matching condition, Eq.~\ref{eq:PM}, it is clear that only the 4WM term $\xi$ creates a dispersive phase shift of pump, signal and idler. In other words, in a pure 3WM case, $\xi=0$ and the phase matching condition becomes $\Delta_k=0$, naturally fulfilled in a dispersion-less transmission line. While detrimental for noise properties (see Sec.~\ref{sec:noise}), we can use the continued presence of 4WM to our advantage, because it allows us to calibrate the pump power, down to the KIT input.
	
	In fact, in such a situation $I_d$ and $I_{p0}$ influence the pump tone phase shift, which we can measure unambiguously (i.e. not $\mod 2\pi$) with a VNA. More precisely, although the phase $\phi=\arg(S_{21})$ read by a VNA is $2\pi$-wrapped, its shift $\delta_\phi = \phi-\phi_0$ from an initial value $\phi_0$ can be continuously monitored when $I_d$ and $I_{p0}$ vary, and thus unambiguously determined. This phase shift in turn translates into a wavenumber variation $\delta_p=-\delta_\phi/N_c$. If initially at zero dc bias and small rf pump amplitude, then $\delta_p=\beta_p - k_p$, with $\beta_p$ the pump wavenumber, dependent on $I_d$ and $I_{p0}$, and $k_p$ the linear wavenumber. When a single pump tone travels along the line (no input signal), we are by default under the strong pump approximation, and the first equation of Eqs.~\ref{eq:CMEstrong} gives $I_p(x)=I_{p0}\exp{(i\xi k_p I_{p0}^2 x /8)}$. In addition, the current I in the line then writes as $I=1/2\{I_p(x)\exp[i(k_px-\omega_pt)] + c.c\}$, and thus the pump wavenumber is $\beta_p=\xi k_p I_{p0}^2/8 + k_p$, which leads to $\delta_p=\xi k_p I_{p0}^2/8$. Because $k_p = \omega_p\sqrt{L_d C}$, we can rewrite:
	\begin{equation}
	\delta_p= \frac{1}{8} \frac{I_{p0}^2}{I_*^2} \omega_p \sqrt{L_0 C} \sqrt{1+\frac{I_d^2}{I_*^2}},
	\end{equation}
	therefore $I_{p0}$ and $I_d$ induce similar phase shift in the line. Knowing $I_d$ (room temperature parameter), we thus get $I_{p0}$ at the KIT input.
	
	\section{ABCD TRANSFER MATRICES}
	\label{app:ABCD}
	The dispersion relations, Fig.~\ref{fig:PMgaintheo}a are calculated by cascading the ABCD matrices of the KIT cells, a method suitable for any periodic loading pattern. We then compute the KIT $S_{21}$ scattering parameter as $S_{21}=2/(A+B/Z_0+CZ_0+D)$ \cite{pozar2011microwave}, where $Z_0=50$ \ohm~is the input and output ports impedance, and finally get $k=-\mathrm{unwrap}[\arg(S_{21})]/N_c$. 
	
	In the unloaded case, represented in Fig.~\ref{fig:schematicKIT}, the ABCD matrix cell is:
	\begin{equation}
	\boldsymbol{T_c} = 
	\begin{bmatrix}
	1 & i L_d \omega\\
	\frac{i2C\omega}{2-L_fC\omega^2} & 1-\frac{2L_dC\omega^2}{2-L_fC\omega^2}
	\end{bmatrix}.
	\end{equation}
	All the cells being identical, the KIT's ABCD matrix is simply $\boldsymbol{T_K} = \boldsymbol{T_c}^{N_c}$. In Fig.~\ref{fig:PMgaintheo}a we used $N_c=6.6\times10^4$ to match the length of our fabricated KIT, and $L_d=45.2$ pH, $C=18.8$ fF, and $L_f=1.02$ nH, values that match our design (fingers are $102$ \micron long and $1$ \micron wide).
	
	In the loaded case, some cells have shorter fingers, see Fig.~\ref{fig:KITfab}a. In these cells, the capacitance to ground is $C_l=L_d/Z_l^2$, where $Z_l$ is the load impedance, and a finger's inductance is $L_l$. To compute the KIT scattering parameter, we form the ABCD matrix of the repetition pattern comprised with unloaded and loaded cells:
	\begin{equation}
	\begin{aligned}
	\boldsymbol{T_\mathrm{sc}} =
	&\begin{bmatrix}
	1 & i L_d \omega\\
	\frac{i2C\omega}{2-L_fC\omega^2} & 1-\frac{2L_dC\omega^2}{2-L_fC\omega^2}
	\end{bmatrix}^{N_u/2}\\
	\times&
	\begin{bmatrix}
	1 & i L_d \omega\\
	\frac{i2C_l\omega}{2-L_lC_l\omega^2} & 1-\frac{2L_dC_l\omega^2}{2-L_lC_l\omega^2}
	\end{bmatrix}^{N_l}\\
	\times&
	\begin{bmatrix}
	1 & i L_d \omega\\
	\frac{i2C\omega}{2-L_fC\omega^2} & 1-\frac{2L_dC\omega^2}{2-L_fC\omega^2}
	\end{bmatrix}^{N_u/2},
	\end{aligned}
	\end{equation}
	where $N_u$ is the number of unloaded cells and $N_l$ is the number of loaded cells in the pattern, which we call a \textit{supercell}. As before, to get the KIT's ABCD matrix, we simply form $\boldsymbol{T_K} = \boldsymbol{T_\mathrm{sc}}^{N_\mathrm{sc}}$, where $N_\mathrm{sc} = N_c/(N_u+N_l)$ is the number of supercells in the KIT. Here, $N_u=60$ (equivalent to $300\upmu$m), $N_l=6$ (equivalent to $30\upmu$m), $N_c=66000$ (equivalent to $33$ cm), therefore $N_\mathrm{sc}=1000$. The plain line in Fig.~\ref{fig:PMgaintheo}a shows the wavenumber $k^*$ thus found, with $Z_l=80$ \ohm, and $L_l=335$ pH, as the finger length in a loaded cell is $33.5$ \micron.
	
	To compute the signal power gain, Fig.~\ref{fig:PMgaintheo}b, we inject the expression of $k(\omega)$ for a periodically loaded KIT (from $\boldsymbol{T_K}$) in the CME, Eqs.~\ref{eq:CME}. We solve them numerically for different pump frequencies. For $8.8812$ GHz (blue curve), $8.8992$ GHz (red), $8.9256$ GHz (green) and $8.9736$ GHz (purple), phase matched signal and idler are detuned from the half pump frequency by $0$, $1$, $1.5$ and $2$ GHz respectively. For $8.855$ GHz (gray curve), phase matching is nowhere achieved. We used $N_c=6.6\times10^4$, $I_*=7$ mA, $I_d=1.5$ mA, and the initial conditions $I_{p0}=I_*/60$, $I_{s0}=I_{p0}/100$, and $I_{i0}=0$, close to experimental values.
	
	\section{KIT SATURATION}
	
	\subsection{Strong signal gain profile asymmetry}
	\label{app:gainasym}
\begin{figure}[h!]
		\centering
		\includegraphics[scale=0.45]{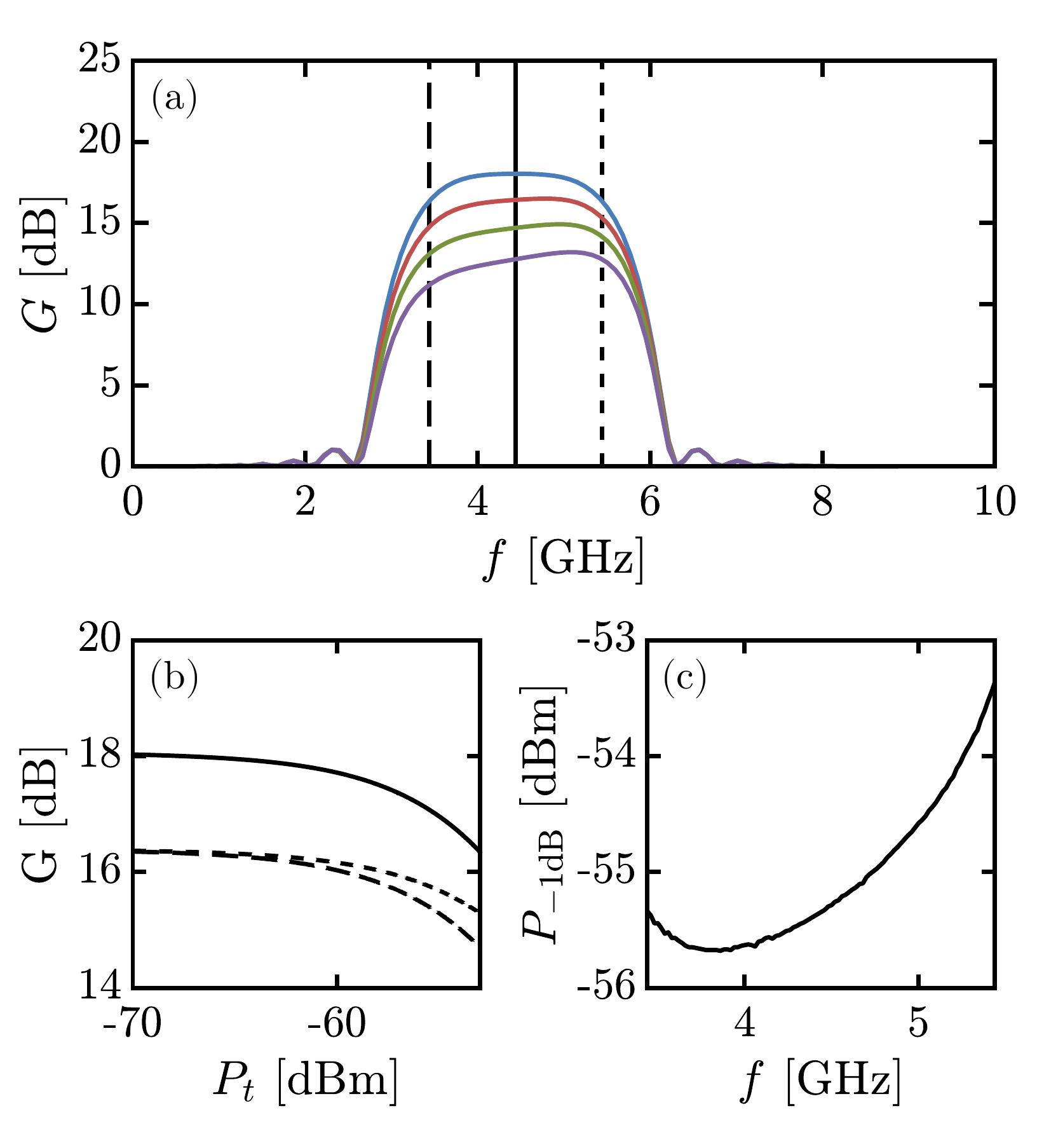}
		\caption{Theory of KIT saturation, calculated by solving the full CME \ref{eq:CME}. (a) Distortion of the gain profile, at four different input signals: $I_{s0}=I_{p0}/100$ (blue curve), corresponding to a small signal regime, $I_{s0}=I_{p0}/12$ (red),  $I_{s0}=I_{p0}/8$ (green), and $I_{s0}=I_{p0}/6$ (purple). The KIT length, is $N_c=6.6\times10^4$ cells. Vertical lines indicate three frequencies: 3.4406 GHz (long dashed), 4.4406 GHz (plain), equal to the half pump frequency, and 5.4406 GHz (short dashed). (b) The gain of a probe tone is calculated at these three frequencies, as a function of the probe tone power. (c) The 1 dB compression power is shown as a function of frequency.}
		\label{fig:satgain}
	\end{figure}

	When the input signal power amounts to a significant fraction of the pump power, parametric amplification depletes the pump. It surprisingly generates asymmetry in the signal gain profile, with respect to the half pump frequency. Figure \ref{fig:satgain}a shows signal gain profiles, calculated when phase matching is achieved at exactly half the pump frequency, i.e. for $\omega_s=\omega_i$ (corresponding to $\omega_p=8.8812$ GHz), at various initial signal powers $P_{s0}$. They are obtained by solving the CME \ref{eq:CME}, which incorporate pump depletion effects. As $P_{s0}$ increases, the gain diminishes, and the originally flat profile tilts, with higher frequencies presenting higher gain.
	
	Fundamentally, this asymmetry stems from the fact that when solving the CME in the case where $I_p$ varies along the KIT transmission line, the initial conditions (ICs) vary as a function of frequency. More precisely, the second and third equations in Eqs.~\ref{eq:CME}, govern the evolution of $I_s$ and $I_i$ respectively; in a simplified version, they write as
	\begin{equation}
	\begin{aligned}
	\frac{d I_s}{dx} &= \frac{i k_s \epsilon}{4} I_p I_i^*\\
	\frac{d I_i}{dx} &= \frac{i k_i \epsilon}{4} I_p I_s^*.
	\end{aligned}
	\label{eq:CMEsimple}
	\end{equation}
	We dropped the second terms in the rhs of Eqs.~\ref{eq:CME}, representing the 4WM conversion processes, as the asymmetry still holds when $\xi=0$, and we assumed perfect phase matching in 3WM, $\Delta_k=0$, i.e. a dispersionless line.
	
	In the undepleted pump regime $I_p(x)=I_{p0}$, and we can decouple the equations on $I_s$ and $I_i$. Deriving with respect to $x$ Eqs.~\ref{eq:CMEsimple}, we get
	\begin{equation}
	\frac{d^2 I_j}{dx^2} = g_3^2 I_j,
	\label{eq:CMEsimplestrong}
	\end{equation}
	with $j\in\{s,i\}$, and $g_3=\frac{\epsilon I_{p0}}{4}\sqrt{k_i k_s}$, as defined in appendix \ref{app:CME}. Using the ICs $I_s(0)=I_{s0}$ and $dI_s/dx(0)=0$ (because $I_i^*(0)=0$), $I_s=\cosh(g_3x)I_{s0}$, as seen in Eqs.~\ref{eq:siPM}. Signal and idler wavenumbers appear as a product in this solution, hence for any $x$ the signal amplitude $I_s$ is symmetric with respect to the half pump frequency.
	
	In the soft pump regime, where $I_p(x)$ is not constant, Eqs.~\ref{eq:CMEsimple} cannot be decoupled. We can however write them in a canonical form, deriving with respect to $x$:
	\begin{equation}
	\frac{d^2 I_j}{dx^2} -\frac{1}{I_p}\frac{d I_p}{dx} \frac{dI_j}{dx} -\frac{k_sk_i\epsilon^2}{16}\lvert I_p \rvert^2 I_j= 0,
	\label{eq:CMEcanon}
	\end{equation}
	with $j\in\{s,i\}$. Here, the interplay between $I_s$ and $I_i$ comes from $I_p$, which contains the product $I_sI_i$ (see Eqs.~\ref{eq:CME}). Although signal and idler wavenumbers also appear only as a product in Eqs.~\ref{eq:CMEcanon}, these CME lead to an asymmetric signal amplitude profile, because out of the five ICs required to solve them, one changes with frequency: $I_p(0)=I_{p0}$, $I_s(0)=I_{s0}$, $I_i(0)=0$, $dI_s/dx(0)=0$, and $dI_i/dx(0)=i\epsilon I_{p0} I_{s0} k_i/4$. This last IC depends on $k_i$, which depends on the signal frequency. In the small signal limit, $dI_i/dx(0)\xrightarrow{}0$, and we recover a symmetric gain profile with respect to the half pump frequency.
	
	\subsection{Compression power calculation}
	\label{app:ph}
	
	This asymmetry produces higher power handling capabilities at frequencies above $\omega_p/2$, compared to below $\omega_p/2$. Thus, considering that only half the bandwidth is usable for resonator readout, it is more advantageous to have these lie above $\omega_p/2$. Figure \ref{fig:satgain}b shows the gain as a function of a probe tone power $P_t$, calculated from the CME \ref{eq:CME} at three frequencies: one at $\omega_p/2$, and two at $\omega_p/2\pm 1$ GHz. Because phase matching is set to be optimal at $\omega_s=\omega_i=\omega_p/2$, gain is maximal at this frequency. The small signal gain is identical for $\omega_p/2\pm 1$ GHz, however it visibly compresses at higher tone power for $\omega_p/2 + 1$ GHz. Figure \ref{fig:satgain}c presents the 1 dB compression power $P_\mathrm{-1dB}$, calculated in the interval $[\omega_p/2-1,\omega_p/2+1]$ gigahertz. As expected, $P_\mathrm{-1dB}$ is a few dB higher when $\omega>\omega_p/2$. This phenomenon is reminiscent of gain distortion, seen in JPAs \cite{malnou2018optimal}. Effects not included in the CME \ref{eq:CME}, such as standing wave patterns, or defects in the line, which locally lower $I_*$, may cause the discrepancy between these theoretical calculations and the measurements (see Sec.~\ref{sec:exp})
	
	\section{KIT PACKAGING}
	\label{app:packaging}
	
	There are three main concerns when packaging a KIT: first, the package should be matched to 50 \ohm. Any mismatch will result in reflections, creating gain ripples (see Sec.~\ref{sec:exp}). Second, the package should ensure good electrical grounding of the transmission line. Otherwise, given the fairly large chip size, spurious chip modes can appear within the frequency range of interest. Third, the package should ensure good thermalization inside the chip. Because the pump power remains high for millikelvin operations, any inhomogeneous rise in temperature may trigger a hot spot near a weak-link, and possibly break superconductivity. We implemented a series of technologies to address these concerns.
	
	\begin{figure}[!h]
		\centering
		\includegraphics[scale=1.3]{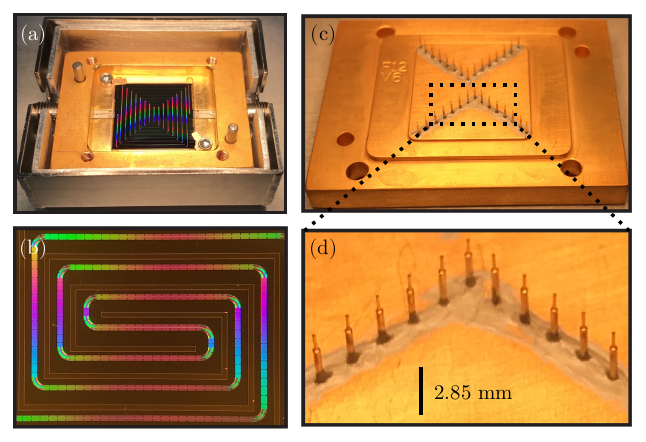} 
		\caption{KIT packaging. (a) The KIT is clamped on a gold plated copper box and wire-bonded to PCBs and to the box itself. Although not very sensitive to magnetic field, the KIT is shielded with aluminum and A4K cryogenic shielding. Two SMA connectors protrude on both side of the packaging. (b) A close-up on the central region of the KIT shows the periodically loaded transmission line, with gold strips deposited between its spiral arms. (c) The top copper lid contains spring-loaded pogo pins inserted half-way into the box. They are arranged to contact the chip between the KIT trace. (d) A close-up on the pins shows their top thinner part, which can retract inside the body of the pin. They are fixed on the copper lid with dried silver paint.} 
		\label{fig:kitpackaging}
	\end{figure}
	Figure \ref{fig:kitpackaging}a presents the chip,
	clamped onto the bottom part of the copper packaging. The chip is wire-bonded on both sides to printed circuit boards (PCBs). They convert the on-chip CPW layout to microstrip, and then the central pin of sub-miniature version A (SMA) connectors are soldered onto the microstrip. We suspect imperfect PCBs, with measured impedance close to 52 \ohm, play a role in creating gain ripples. When designing the spiral, we carefully adjusted the radius of the turns, in conjunction with the unit cell length, to have these turns match the straight sections' inductance and capacitance per length.
	
	Electrical grounding inside the chip is ensured with pogo pins inserted in the top lid of the packaging, see Fig.~\ref{fig:kitpackaging}c and d. When closing the box, these pins contact the chip between the line traces. If absent, we have measured spurious resonant modes with harmonics at gigahertz frequencies. Pins are $140$ \micron in diameter, and each applies a 20 g force to the chip.
	
	\begin{figure*}[t] 
		\centering
		\includegraphics[scale=0.61]{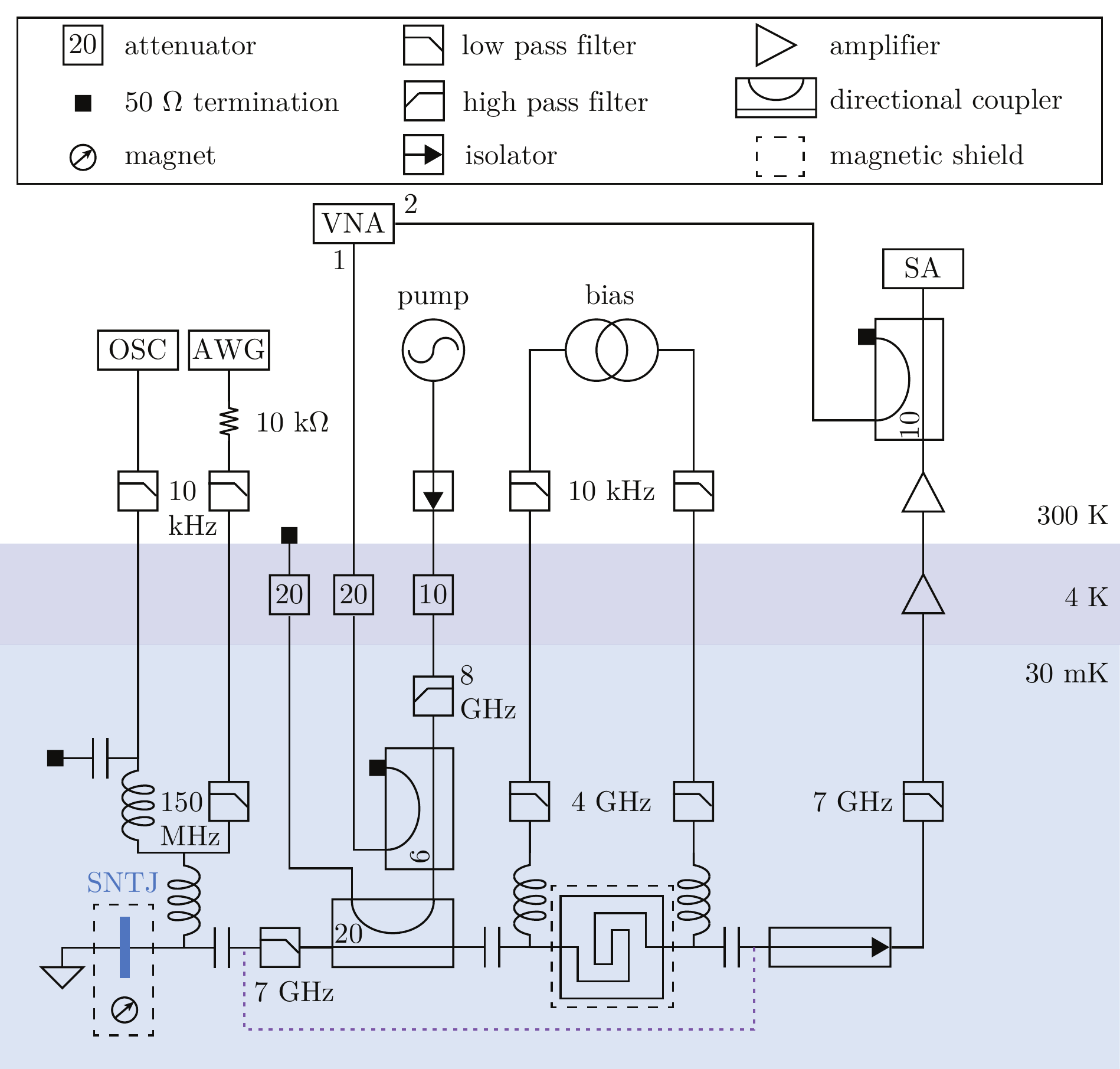}
		\caption{Full schematic of the noise measurement experimental setup. The KIT is represented by a spiral, enclosed in a square. A permanent magnet suppresses the Josephson effect in the SNTJ. A magnetic shield protects other microwave components from the effect of this magnet. The 4-12 GHz isolator is from Quinstar technology, model CWJ1015-K13B. The dashed purple line represents the bypass.}
		\label{fig:noisefullsetup}
	\end{figure*}
	These pins also act as thermal links to the packaging. In addition, we deposited gold strips onto the NbTiN layer, inside the spiral, between the KIT line traces, and near the chip edges, see Fig.~\ref{fig:kitpackaging}b. These strips contact the pins. Absent the pogo pins, superconductivity breaks down before high gain can be reached. Finally, we gold-bond the chip ground plane (from the deposited gold) to the copper box (instead of using standard aluminum bonding): gold remains a normal metal at millikelvin temperatures, thereby better thermalizing the chip.
	
	\section{NOISE MEASUREMENT EXPERIMENTAL SETUP}
	\label{app:setupnoise}
	
	Figure \ref{fig:noisefullsetup} presents a schematic of the full experimental setup used to measure the system-added noise of a readout chain using a KIT as a pre-amplifier. In total, noise generated by the SNTJ travels through three amplifiers: the KIT, a HEMT at 4K, and a room temperature amplifier, before finally being recorded with a SA.
	
	The SNTJ is packaged with a permanent magnet suppressing the Josephson effect, and a magnetic shield protects other elements from this magnetic field. In addition, a bias tee routes SNTJ-generated rf noise to the microwave readout chain, while at the same time allowing for dc bias. In fact, the SNTJ is biased with an arbitrary waveform generator (AWG). It outputs a low frequency (50 Hz) triangular voltage wave on a $10$ k\ohm~current limiting resistor, thereby creating a current $I_\mathrm{AWG}$, varying between $\pm12$\microamp, which sweeps the SNTJ-generated noise value. 
	
	An oscilloscope reads the SNTJ voltage \textit{in situ} while the AWG outputs a known current, allowing the computation of the SNTJ impedance $Z_\mathrm{SNTJ}=54\pm4$ \ohm, and with it, the SNTJ voltage bias $V=Z_\mathrm{SNTJ}I_\mathrm{AWG}$.
	
	Noise from the SNTJ is combined with rf tones (pump, and probe from a VNA to measure the gain profile) via a 20 dB directional coupler (DC) connected to the KIT input. Additionally, a 7 GHz low-pass filter placed between the SNTJ and the DC prevents the rf pump from leaking back to the SNTJ. Because the KIT requires a fairly high pump power ($-29$ dBm), we only attenuate the pump line by 10 dB at 4K. Then, an 8 GHz high-pass filter at 30 mK rejects noise at frequencies within the KIT amplification band, while allowing the pump to pass. A bias tee at the KIT input port combines rf signals (including noise from the SNTJ) with the KIT dc current bias, and a second bias tee at the KIT output separates dc from rf. The rf signal then passes through a $4-12$ GHz isolator, and a 7 GHz low-pass filter (Pasternack PE87FL1015, $\sim45$ dB rejection at 9 GHz), preventing the pump tone from saturating the HEMT.
	
	The SA is operated in a zero-span mode, its acquisition triggered by the AWG. This measures the output noise at a single frequency, over a 5 MHz resolution bandwidth (RBW), and directly traces out the curves of Fig.~\ref{fig:KITnoise}b. Varying the SA center frequency, we obtain the system-added noise over the full 3-6.5 GHz bandwidth, Fig.~\ref{fig:KITnoise}c.
	
	\section{NOISE THEORY}
	
	\subsection{System-added noise}
	\label{app:noise}
	
	When propagating through the experimental setup presented in Fig.~\ref{fig:noisefullsetup}, noise generated by the SNTJ undergoes loss and amplification. Both processes affect the effective noise at each amplifier input, and therefore also, the noise reaching the SA. While the overall system-added noise $N_\Sigma$ encompasses microwave loss, we derive its complete expression as a function of signal ($N_\mathrm{ex}^s$) and idler ($N_\mathrm{ex}^i$) amplifier-excess noise, gain, and transmission efficiencies to estimate $N_\mathrm{ex}^s+N_\mathrm{ex}^i$.
	
	Figure \ref{fig:setupnoise} represents the lossy amplification chain, with the transmission efficiencies, gain and added noise associated with each interconnection and amplification stages. We have:
	\begin{align}
	N_1^{j} &= \eta_1^j \left[N_\mathrm{in}^j + \frac{N_f(1-\eta_1^j)}{\eta_1^j}\right] \label{eq:N1}\\
	N_2^s &= G(N_1^s + N_\mathrm{ex}^s) + (G-1)(N_1^i + N_\mathrm{ex}^i) \label{eq:N2}\\
	N_3^s &= \eta_2 \left[N_2^s + \frac{N_f(1-\eta_2)}{\eta_2}\right] \label{eq:N3}\\
	N_4^s &= G_H(N_3^s + N_H) \label{eq:N4}\\
	N_o^s &= G_r N_4^s \label{eq:Nos_app_basic}.
	\end{align}
	Here, $N_1^j$ with $j\in\{s,i\}$ is the KIT-input noise at the signal and idler frequency respectively; then, at the signal frequency, $N_2^s$ is the KIT-output noise, $N_3^s$ is the HEMT-input noise, $N_4^s$ is the HEMT-output noise, and $N_o^s$ is the noise measured by the SA. Note that the beamsplitter interaction between the KIT and the HEMT assumes the loss to be mostly cold, which is the case in our setup where the lossy components before the HEMT are at the 30 mK stage. Refer to table \ref{tab:variables} for the other variable definitions. From Eqs.~\ref{eq:N1}-\ref{eq:Nos_app_basic} we can derive:
	\begin{align}
	N_o^s &= G_c^{ss}(N_\mathrm{in}^s + N_\mathrm{eff}^s) + G_c^{si}(N_\mathrm{in}^i + N_\mathrm{eff}^i) \label{eq:Nos_app} \\
	& = G_c^{ss}\left[N_\mathrm{in}^s + N_\mathrm{eff}^s + \frac{G-1}{G}\frac{\eta_1^i}{\eta_1^s}(N_\mathrm{in}^i+N_\mathrm{eff}^i)\right] \label{eq:Nos_app_detail},
	\end{align}
	where 
	\begin{align}
	G_c^{ss} &=G_r G_H\eta_2 G\eta_1^s \label{eq:Gcss}\\
	G_c^{si} &=G_r G_H\eta_2 (G-1)\eta_1^i \label{eq:Gcsi},
	\end{align}
	and where
	\begin{align}
	N_\mathrm{eff}^s &= \frac{N_\mathrm{ex}^s + (1-\eta_1^s)N_f}{\eta_1^s} + \frac{(1-\eta_2)N_f+N_H}{\eta_2 G\eta_1^s} \label{eq:Neffsapp} \\
	N_\mathrm{eff}^i &= \frac{N_\mathrm{ex}^i + (1-\eta_1^i)N_f}{\eta_1^i} \label{eq:Neffiapp}.
	\end{align}
	The system-added noise is defined as the part of the output noise in Eq.~\ref{eq:Nos_app_detail} that is not due to the input $N_\mathrm{in}^s$, and by assuming a cold input at the idler port, i.e. assuming $N_f=N_\mathrm{in}^i$. We thus find
	\begin{equation}
	N_\Sigma = N_\mathrm{eff}^s + \frac{G-1}{G}\frac{\eta_1^i}{\eta_1^s}(N_f+N_\mathrm{eff}^i).
	\label{eq:Nsigma_app}
	\end{equation}
	This equation is equivalent to Eq.~\ref{eq:Nsigma}, where we did not simplify the ratio $G_c^{si}/G_c^{ss}$.
	
	\begin{figure*}[t] 
		\centering
		\includegraphics[scale=0.6]{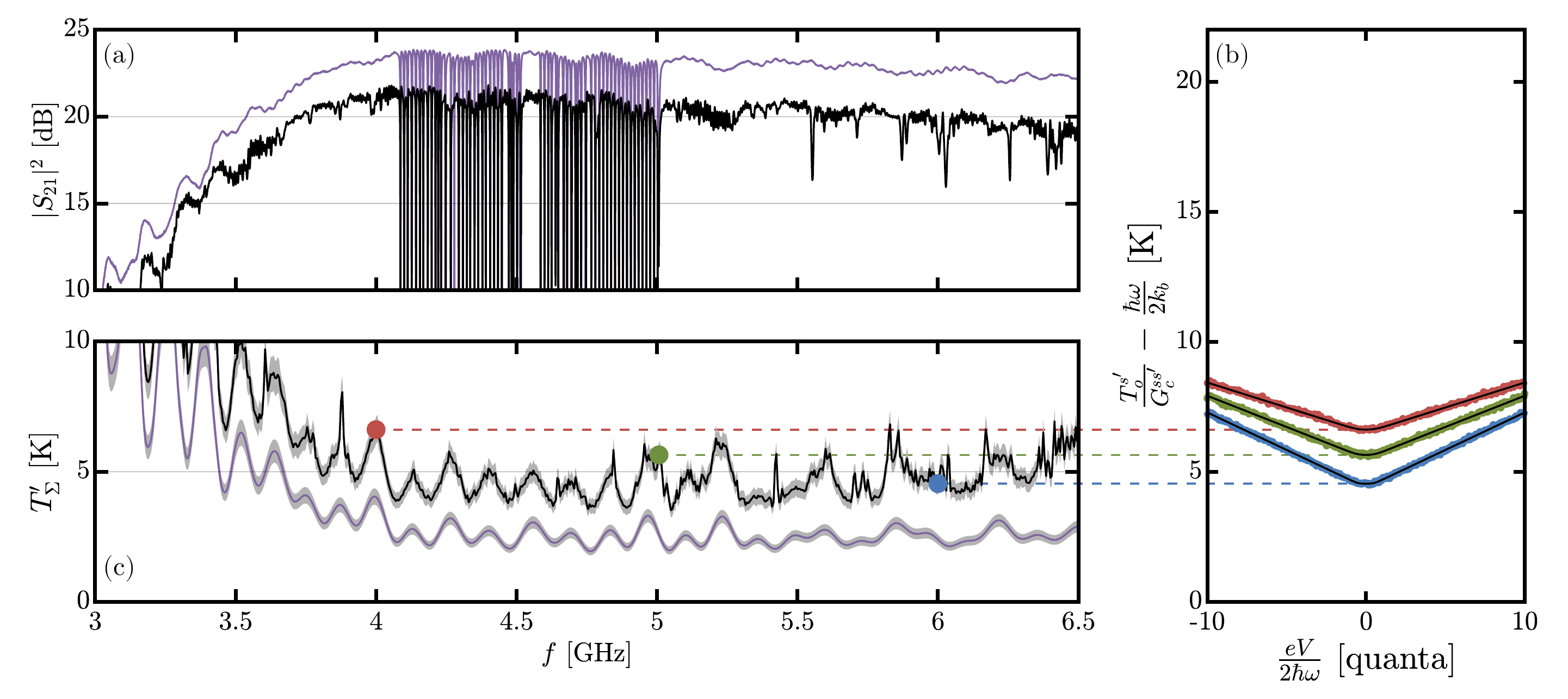} 
		\caption{System-added noise measurement of a microwave amplification chain with the HEMT as first amplifier. (a) The transmission through the whole noise setup (KIT un-pumped), when the SNTJ has been replaced by a transmission line capacitively coupled to an array of resonators (not used in the present experiment). It is performed in two situations: without (black curve) and with (purple curve) the bypass. (b) Examples of $T_o^{s'} = N_o^{s'}\hbar\omega/k_B$ are shown for the SA centered at 4, 5 and 6 GHz, with fits superimposed in black lines. The SA RBW is 5 MHz. (c) From the fits, we extract the system-added noise temperature $T_\Sigma' = N_\Sigma'\hbar\omega/k_B$ as a function of frequency without (black curve) and with (purple curve) the bypass. Uncertainties are indicated by the gray area surrounding the lines.} 
		\label{fig:HEMTnoise}
	\end{figure*}
	
	Assuming $\{G,N_H\}\gg1$ and inserting Eqs.~\ref{eq:Neffsapp} and \ref{eq:Neffiapp} into Eq.~\ref{eq:Nsigma_app} we get
	\begin{equation}
	N_\Sigma = \frac{N_\mathrm{ex}^s+N_\mathrm{ex}^i}{\eta_1^s} + \frac{2(1-\eta_1^s)N_f}{\eta_1^s} + \frac{N_H}{\eta_2 G \eta_1^s} + N_f.
	\end{equation}
	which is Eq.~\ref{eq:Nsigma_simple} presented in the main text.
	
	When the KIT is not pumped, we consider it as a lossless, noiseless, passive element. We therefore have $G=1$, $N_\mathrm{ex}^s=0$ and $N_\mathrm{ex}^i=0$. In that situation
	\begin{equation}
	N_o^{s'} = G_c^{ss'} (N_\mathrm{in}^s + N_\Sigma'),
	\label{eq:Nosp}
	\end{equation}
	where $G_c^{ss'} = G_r G_H\eta_2 \eta_1^s$. From Eq.~\ref{eq:Nsigma_app} and \ref{eq:Neffsapp} we thus get
	\begin{equation}
	N_\Sigma' = \frac{(1-\eta_2\eta_1^s)N_f+N_H}{\eta_2\eta_1^s}
	\label{eq:Nsp}
	\end{equation}
	as the chain's system-added noise with the HEMT as first amplifier.
	
	\subsection{Discarding the idler port input noise}
	\label{app:noisecorrection}
	In a simple case where $G-1\simeq G$, and $\eta_1^s\simeq\eta_1^i$, Eq.~\ref{eq:Nos_app} becomes
	\begin{equation}
	N_o^s = G_c^{ss}(N_\mathrm{in}^s + N_\mathrm{in}^i + N_\mathrm{eff}^s + N_\mathrm{eff}^i).
	\label{eq:Nossimple}
	\end{equation}
	Thus, varying the SNTJ bias i.e. $N_\mathrm{in}^s$ and $N_\mathrm{in}^i$ simultaneously, the y-intercept gives $N_\mathrm{eff}^s + N_\mathrm{eff}^i$, equal to zero for a quantum-limited amplifier. Also, in that simple case, the system-added noise is $N_\Sigma = N_f + N_\mathrm{eff}^s + N_\mathrm{eff}^i$, as shown by Eq.~\ref{eq:Nsigma_app}.
	
	On the other hand, if the calibrated noise (coming from the SNTJ or any other wideband noise source, such as a hot/cold load, or a variable temperature stage) only illuminates the signal port of the amplifier, or if  the  noise at the idler port is wrongly discarded from the analysis, we get $N_\mathrm{in}^i = N_f$ and Eq.~\ref{eq:Nos_app} becomes
	\begin{equation}
	N_o^s = G_c^{ss}(N_\mathrm{in}^s + N_f + N_\mathrm{eff}^s + N_\mathrm{eff}^i).
	\label{eq:Noswrong}
	\end{equation}
	Usually, no distinction is made between $N_\mathrm{eff}^s$ and $N_\mathrm{eff}^i$, and the effective excess noise is simply $N_\mathrm{eff} = N_\mathrm{eff}^s + N_\mathrm{eff}^i$. The sum $N_f + N_\mathrm{eff}$ is then what is commonly defined as the system-added noise.	Varying the SNTJ bias, i.e. $N_\mathrm{in}^s$, the y-intercept gives $N_f + N_\mathrm{eff}$, equal to half a quantum for a quantum-limited amplifier. 
	
	In practice, we fit Eqs.~\ref{eq:Nossimple} or \ref{eq:Noswrong} to find the chain's gain and the y-intercept. Fitting a situation correctly described by Eq.~\ref{eq:Nossimple} with Eq.~\ref{eq:Noswrong} leads to an underestimate of the system-added noise. In fact, assuming $N_\mathrm{in}^s \simeq N_\mathrm{in}^i$, Eq.~\ref{eq:Nossimple} can be rewritten into a form comparable to Eq.~\ref{eq:Noswrong}:
	\begin{equation}
	N_o^s \simeq 2G_c^{ss}\left(N_\mathrm{in}^s + \frac{N_\mathrm{eff}}{2}\right).
	\end{equation}
	The \textit{interpretation} of the y-intercept is crucial. Here, the fit yields $N_\mathrm{eff}/2$ as the y-intercept, and $2G_c^{ss}$ as the chain's gain. The true system-added noise is then twice the y-intercept value plus a half quantum, $N_\Sigma = N_\mathrm{eff} + N_f$, and the true chain's gain is half of the fitted slope. Conversely, assuming Eq.~\ref{eq:Noswrong} leads to the conclusion that the y-intercept value is already the system-added noise, and that the slope is the chain's gain. The true system-added noise is thus more than twice as high, and the chain's true gain is half as much (i.e. 3 dB lower, a mistake usually hard to detect).
	
	\subsection{Shot-noise fit}
	\label{app:sn}
	
	The SNTJ generates a known noise power \cite{spietz2006shot}. The amount of noise $N_\mathrm{in}^j$, with $j\in\{s,i\}$ delivered to the 50 \ohm~transmission line can be written as \cite{lecocq2017nonreciprocal}
	\begin{equation}
	\begin{aligned}
	N_\mathrm{in}^j = \frac{k_B T}{2 \hbar\omega_j}\bigg[&\frac{eV+\hbar\omega_j}{2k_B      T}\coth\left(\frac{eV+\hbar\omega_j}{2k_B T}\right)\\
	+ &\frac{eV-\hbar\omega_j}{2k_B T}\coth\left(\frac{eV-\hbar\omega_j}{2k_B T}\right)\bigg],
	\end{aligned}
	\label{eq:Ni}
	\end{equation}
	where T is the physical temperature of the SNTJ, and V the SNTJ voltage bias. In practice, the AWG has a slight voltage offset, which we include as a fit parameter: we write $V-V_\mathrm{off}$ instead of V in Eq.~\ref{eq:Ni}. In a first step we fit the asymptotes of the output noise response, for which $\lvert eV/(2\hbar\omega) \rvert > 3$ quanta. In that case, Eq.~\ref{eq:Ni} reduces to $N_\mathrm{in}^j = eV/(2\hbar\omega_j)$, and thus Eq.~\ref{eq:Nos_app} reduces to
	\begin{equation}
	\begin{aligned}
	N_o^s = &G_c^{ss}\left(\frac{eV-V_\mathrm{off}}{2\hbar\omega_s} + N_\mathrm{eff}^s\right) \\ + &G_c^{si}\left(\frac{eV-V_\mathrm{off}}{2\hbar\omega_i} + N_\mathrm{eff}^i \right).
	\end{aligned}
	\end{equation}
	We thus get $V_\mathrm{off}$,  $G_c^{ss}$ and $G_c^{si}$. In a second step, we fix $V_\mathrm{off}$,  $G_c^{ss}$ and $G_c^{si}$ to their values derived in the first step and fit the central region (where $\lvert eV/(2\hbar\omega) \rvert \leq 3$ quanta) to Eq.~\ref{eq:Nos_app} to get $N_\mathrm{eff}^s$, $N_\mathrm{eff}^i$ and T.
	
	The spectrum analyzer (SA) records a power spectrum $P_o^s$ (in Watts), which we convert into a number of photons: $N_o^s = P_o^s/(B\hbar\omega)$, where B is the SA resolution bandwidth. Dividing by $G_c^{ss}$, we then refer $N_o^s$ to the chain's input. Finally, we can subtract $N_f$ (in quanta), to visually read $N_\Sigma$ at $V=0$ on the SNTJ curves of Fig.~\ref{fig:KITnoise}b. In fact, 
	\begin{equation}
	\frac{N_o^s(0)}{G_c^{ss}} - N_\mathrm{in}^s(0) - \frac{G_c^{si}}{G_c^{ss}}N_\mathrm{in}^i(0) + N_f\frac{G_c^{si}}{G_c^{ss}} = N_\Sigma.
	\label{eq:Nosrefinsubtract}
	\end{equation}
	At $V=0$,
	\begin{equation}
	N_\mathrm{in}^j(0) = \frac{1}{2}\coth{\left(\frac{\hbar\omega_j}{k_BT}\right)},
	\end{equation}
	where $j\in\{s,i\}$, therefore if $\hbar\omega_j \gg k_BT$, $ N_\mathrm{in}^j(0)\simeq N_f = 0.5$. Considering that $G_c^{ss}\simeq G_c^{si}$, Eq.~\ref{eq:Nosrefinsubtract} yields:
	\begin{equation}
	\frac{N_o^s(0)}{G_c^{ss}} - N_f \simeq N_\Sigma.
	\end{equation}
	
	\subsection{List of variables}
	Table \ref{tab:variables} lists the variables used throughout Sec.~\ref{sec:noise} and appendix \ref{app:noise}.
	
	\begin{table*}[] 
		\centering
		\begin{tabular}{cl}
			\hline
			\hline
			\textbf{variable name} & \textbf{definition} \\ \hline\hline
			$N_\mathrm{in}^s$ & SNTJ-generated noise at the signal frequency\\
			$N_\mathrm{in}^i$ & SNTJ-generated noise at the idler frequency\\
			$N_f$ & vacuum (or thermal) noise, set by the refrigerator temperature\\
			$N_\mathrm{ex}^s$ & signal-to-signal path KIT-excess noise\\
			$N_\mathrm{ex}^i$ & idler-to-signal path KIT-excess noise\\
			$N_\mathrm{eff}^s$ & signal-to-signal path effective KIT-excess noise, Eq.~\ref{eq:Neffsapp}\\
			$N_\mathrm{eff}^i$ & idler-to-signal path effective KIT-excess noise, Eq.~\ref{eq:Neffiapp}\\
			$N_H$ & HEMT-added noise\\
			$N_o^s$ & noise measured by the SA at the signal frequency, Eq.~\ref{eq:Nos_app}\\
			$N_o^{s'}$ & noise measured by the SA when the KIT is off, Eq.~\ref{eq:Nosp}\\
			$N_\Sigma$ & system-added noise, Eq.~\ref{eq:Nsigma_app}\\
			$N_\Sigma'$ & system-added noise when the KIT is off, Eq.~\ref{eq:Nsp}\\
			$\eta_1^s$ & transmission efficiency between SNTJ and KIT at the signal frequency\\
			$\eta_1^i$ & transmission efficiency between SNTJ and KIT at the idler frequency\\
			$\eta_2$ & transmission efficiency between KIT and HEMT at the signal frequency\\
			$G$ & KIT signal power gain\\
			$G_H$ & HEMT signal power gain\\
			$G_r$ & room-temperature signal power gain\\
			$G_c^{ss}$ & signal-to-signal chain's gain, Eq.~\ref{eq:Gcss}\\
			$G_c^{si}$ & idler-to-signal chain's gain, Eq.~\ref{eq:Gcsi}\\
			\hline
		\end{tabular} 
		\caption{List of the variables used in the noise theory. All the variables designating a noise quantity are in units of quanta. All the transmission efficiencies are dimensionless. All the gains are in linear units.}
		\label{tab:variables}
	\end{table*}

	\section{LOSS BUDGET IN THE NOISE MEASUREMENT SETUP}
	\label{app:lossbudget}
	
	To quote the KIT-excess noise terms $N_\mathrm{ex}^s$ and $N_\mathrm{ex}^i$, it is necessary to account for the transmission efficiencies (and therefore the loss) in the measurement setup, $\eta_1^s$, $\eta_1^i$ and $\eta_2$. Note that $\eta_1^i$ is simply the transmission efficiency at the idler frequency $\omega_i=\omega_p-\omega_s$, symmetric of the signal frequency $\omega_s$ with respect to the half-pump frequency $\omega_p/2$. Thus, a broadband measurement of $\eta_1^s$ on both sides of $\omega_p/2$ provides both $\eta_1^s$ and $\eta_1^i$. We estimate these efficiencies in this appendix and give an overall loss budget per component between the SNTJ and the HEMT. Knowing where the loss comes from also provides guidance on how to improve the amplifier because many lossy components could be optimized, particularly by integrating them on-chip with the KIT.
	
	\subsection{System-added noise temperature with un-pumped KIT}
	\label{app:noiseHEMT}
	
	First, we measured the system-added noise temperature of the amplification chain $T_\Sigma'=N_\Sigma' \hbar\omega/k_B$ (with $\hbar$ the reduced Planck's constant and $k_B$ the Boltzmann's constant) when the KIT is off, i.e. with the HEMT as the first amplifier. In that case, $N_\Sigma'$ is given by Eq.~\ref{eq:Nsp}. We measured $N_\Sigma'$ in two situations: one for the measurement setup presented in Fig.~\ref{fig:noisefullsetup}, and one where the KIT surrounded by its two bias tees, the directional coupler and the low pass filter next to it have been bypassed, i.e. replaced by a microwave cable. In the limit where $N_f\ll N_\Sigma'$, the ratio in $N_\Sigma'$ between these two situations gives direct access to the bypassed components' insertion loss (IL) $\mathcal{I}_\mathrm{BP}$ (the exact transmission efficiency ratio from which we calculate $\mathcal{I}_\mathrm{BP}$ is equivalent to finding $\eta_2\eta_1^s$ in Eq.~\ref{eq:Nsp}). Figure \ref{fig:HEMTnoise}c shows $T_\Sigma'$ as a function of frequency, obtained from fitting curves like those presented in Fig.~\ref{fig:HEMTnoise}b, obtained with and without the bypass. There are ripples with 130 MHz characteristic frequency, likely due to reflections in coaxial cables between the SNTJ and the HEMT. Without the bypass $T_\Sigma' = 5.1\pm1.4$ K between 3.5 and 5.5 GHz, while with the bypass $T_\Sigma' = 2.9\pm1$ K. Thus, the ratio gives $\mathcal{I}_\mathrm{BP}=2.5$ dB.
	
	\subsection{Component loss}
	
	Second, we replaced the SNTJ (including the bias tee) with a transmission line, to measure the transmission of a probe tone through the setup (with a VNA). This transmission line is capacitively coupled to an array of resonators, whose resonant frequencies span between $4$ and $5$ GHz, and is part of a future experiment. These resonances are not relevant for the current characterization. We measured the transmission with and without the bypass, shown in Fig.~\ref{fig:HEMTnoise}a. Once again, the transmission ratio between these two situations gives direct access to the bypassed components' IL. We find $\mathcal{I}_{BP} = 2.4\pm0.6$ dB, in agreement with the system-added noise measurements of Fig.~\ref{fig:HEMTnoise}c.
	
	\begin{table*}[] 
		\centering
		\begin{tabular}{cccccccc}
			\hline
			\hline
			\textbf{authors} & \textbf{mixing} & \textbf{\begin{tabular}[c]{@{}c@{}}pump\\ power {[}dBm{]}\end{tabular}} & \textbf{\begin{tabular}[c]{@{}c@{}}gain \\  {[}dBm{]}\end{tabular}} & \textbf{\begin{tabular}[c]{@{}c@{}}bandwidth \\ {[}GHz{]}\end{tabular}} & \textbf{\begin{tabular}[c]{@{}c@{}}saturation\\{[}dBm{]}\end{tabular}} & \textbf{\begin{tabular}[c]{@{}c@{}}noise\\ bandwidth\end{tabular}} & \textbf{\begin{tabular}[c]{@{}c@{}}system\\ {noise [}K{]}\end{tabular}} \\ \hline\hline
			this work & 3WM & $-29$ & $16.5^{+1}_{-1.3}$ & 2 & $-63$ & $3$-$6.5$ GHz& $0.66\pm0.15$\\ 
			Eom et.al. \cite{eom2012wideband} & 4WM & $-8$ & $10^{+3}_{-3}$ & $4$ & $-52$ & $1$ Hz & $1.5$\footnote{\label{ftn:ftn2}idler noise input not accounted for}\\
			Bockstiegel et.al. \cite{bockstiegel2014development} & 4WM & $-10$ & $20^{+3}_{-3}$ & $8$ & - & - & - \\
			Vissers et.al. \cite{vissers2016low} & 3WM & $-10$ & $20^{+5}_{-5}$ & $4$ & $-45$\footnote{at $10$ dB gain} & - & - \\
			Chaudhuri et.al. \cite{chaudhuri2017broadband} & 4WM & $-10$ & $15^{+3}_{-3}$ & $3$ & - & - & - \\
			Ranzani et.al. \cite{ranzani2018kinetic} & 3WM & $-30$ & $10^{+2}_{-2}$ & $4$ & - & $6$-$10$ GHz & $1.5$-$5^\mathrm{\ref{ftn:ftn2}}$ \\
			Zobrist et.al. \cite{zobrist2019wide} & 3WM & $-23$ & $15$\footnote{\label{ftn:ftn1}data quoted but not shown in the paper} & $5^\mathrm{\ref{ftn:ftn1}}$ & $-53^\mathrm{\ref{ftn:ftn1}}$ & 1 MHz & ${0.58^{+0.2}_{-0.03}}^\mathrm{\ref{ftn:ftn2}}$ \\ \hline
		\end{tabular} 
		\caption{Comparison of our KIT to other published KIT results. As discussed in appendix \ref{app:noisecorrection}, failure to account for the noise at the idler input results in an underestimate of the system-added noise temperature by about a factor of two.}
		\label{tab:kittable}
	\end{table*}
	
\begin{figure}[!h]
		\centering
		\includegraphics[scale=0.45]{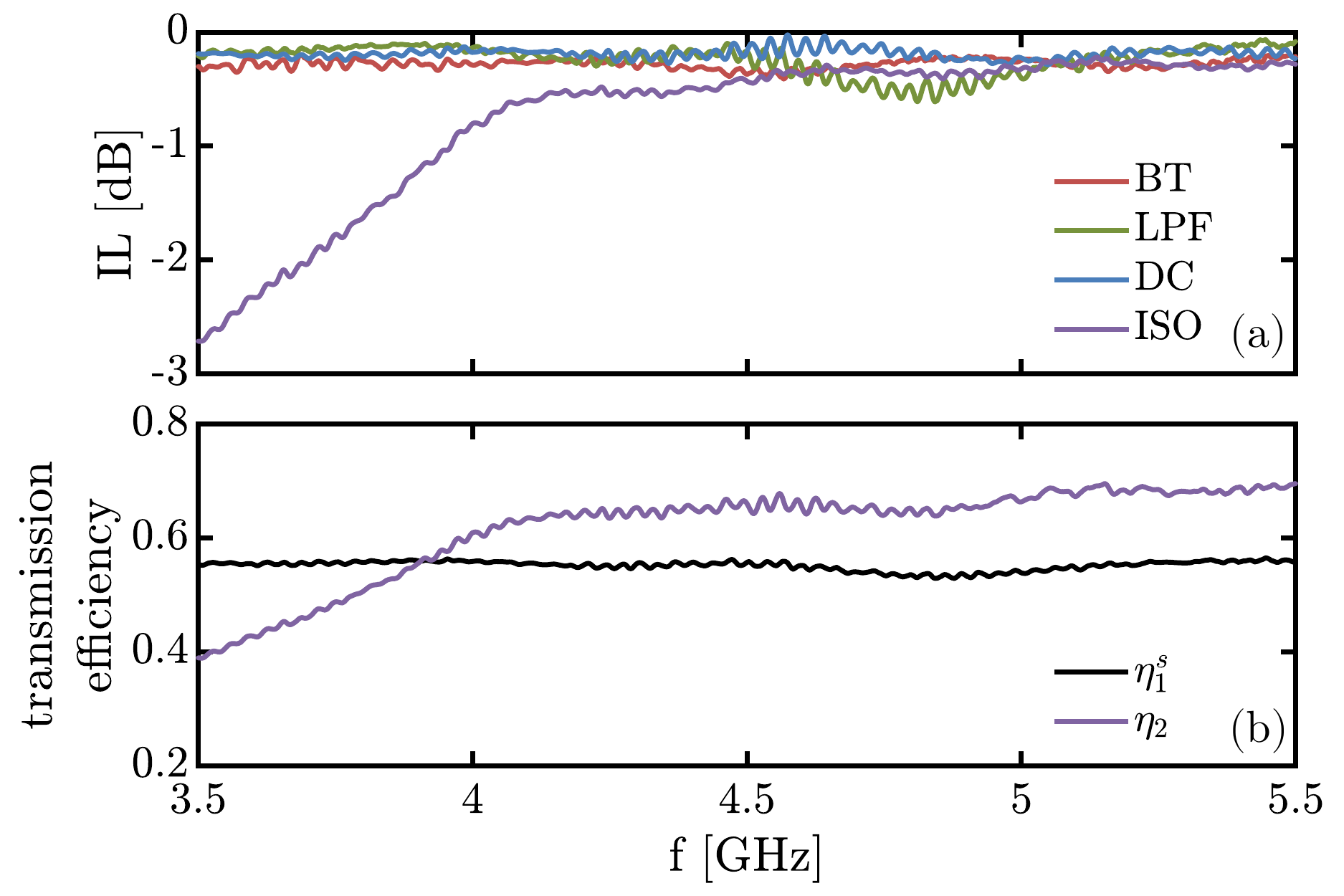} 
		\caption{Insertion loss (IL) from the microwave components used in the noise measurement setup (see Fig.~\ref{fig:noisefullsetup}). The ILs (a) are measured at 4K: the bias tee, Anritsu K250 (red curve), the low pass filter, Pasternack PE87FL1015 (green curve), the directional coupler, Pasternack PE2204-20 (blue curve), and the isolator, Quinstar CWJ1015-K13B (purple curve). (b) Combining the IL from the components before and after the KIT, we estimate the transmission efficiencies $\eta_1^s$ (black curve) and $\eta_2$ (purple curve) respectively.}
		\label{fig:componentsloss}
	\end{figure}

	We also measured the individual transmissions of the chain's components at 4K: bias tee (BT), filter (LPF), directional coupler (DC) and isolator (ISO). Figure \ref{fig:componentsloss}a shows the IL of these four components. Between $3.5$ and $5.5$ GHz, $\mathcal{I}_\mathrm{BT} = 0.3\pm0.04$ dB, $\mathcal{I}_\mathrm{LPF} = 0.2\pm0.1$ dB, $\mathcal{I}_\mathrm{DC} = 0.2\pm0.04$ dB, and $\mathcal{I}_\mathrm{ISO} = 0.7\pm0.6$ dB. Its IL increases below $4$ GHz as we leave its operating band, which degrades the system-added noise performance in the same manner as reducing the KIT gain would, see Eq.~\ref{eq:Neffsapp}. The SNTJ packaging loss, including the bias tee, has been previously reported to be $\mathcal{I}_\mathrm{SNTJ} = 1$ dB (transmission efficiency of $0.8$) \cite{chang2016noise}.
	
	\subsection{HEMT-added noise, transmission efficiencies, KIT-excess noise}
	
	We can estimate the HEMT-added noise temperature $T_H=N_H\hbar\omega/k_B$ from eq.~\ref{eq:Nsp}, because we measured $T_\Sigma'$ (see Fig.~\ref{fig:HEMTnoise}c), and because we can have an estimation of the total IL $\mathcal{I}_T$ from the SNTJ to the HEMT. In fact, without the bypass
	\begin{equation}
	\mathcal{I}_T=\mathcal{I}_\mathrm{SNTJ}+\mathcal{I}_\mathrm{BP}+\mathcal{I}_\mathrm{ISO}+\mathcal{I}_\mathrm{LPF}
	\end{equation}
	(in dB), which gives $\mathcal{I}_T=4.3\pm0.6$ dB. Then, $\eta_2\eta_1^s = 10^{-\mathcal{I}_T/10}$, and we get $T_H=1.8\pm0.2$ K (i.e. $N_H=8\pm1$ quanta) between 3.5 and 5.5 GHz, in agreement with the HEMT data sheet, which gives $T_H=1.6\pm0.3$~K in that frequency range.
	
	Subtracting $\mathcal{I}_\mathrm{LPF}$, $\mathcal{I}_\mathrm{DC}$ and $2\times\mathcal{I}_\mathrm{BT}$ from $\mathcal{I}_\mathrm{BP}$, the KIT's packaging (see Fig.~\ref{fig:kitpackaging}a) is responsible for about $\mathcal{I}_\mathrm{KIT}=1.4\pm0.6$ dB of IL. This loss may be decreased in future optimization, for example by coating the PCBs and the SMA pins with superconducting material.
	
	Finally, we can separately estimate $\eta_1^s$ and $\eta_2$ by adding (in dB) the IL of the chain's components, and then use Eqs.~\ref{eq:Neffsapp} and \ref{eq:Neffiapp} to estimate the KIT-excess noise $N_\mathrm{ex}^s$ and $N_\mathrm{ex}^i$. Between the SNTJ and the KIT,
	\begin{equation}
	\mathcal{I}_{\eta_1^s}=\mathcal{I}_\mathrm{SNTJ}+\mathcal{I}_\mathrm{LPF}+\mathcal{I}_\mathrm{DC}+\mathcal{I}_\mathrm{BT}+\frac{\mathcal{I}_\mathrm{KIT}}{2},
	\end{equation}
	which gives $\mathcal{I}_{\eta_1}=2.4\pm0.1$ dB, and between the KIT and the HEMT
	\begin{equation}
	\mathcal{I}_{\eta_2}=\frac{\mathcal{I}_\mathrm{KIT}}{2}+\mathcal{I}_\mathrm{BT}+\mathcal{I}_\mathrm{ISO}+\mathcal{I}_\mathrm{LPF},
	\end{equation}
	which gives $\mathcal{I}_{\eta_2}=1.9\pm0.6$ dB. Figure \ref{fig:componentsloss}b shows $\eta_1^s=10^{-\mathcal{I}_{\eta_1}/10}$ and $\eta_2=10^{-\mathcal{I}_{\eta_2}/10}$ thus obtained. Therefore, between 3.5 and 5.5 GHz, $\eta_1^s=\eta_1^i=0.57\pm0.02$ (both equal since the frequency band is nearly symmetric with respect to $\omega_p/2=4.4$ GHz), and $\eta_2=0.64\pm0.10$. 
	
	\section{KIT PERFORMANCE COMPARISON}
	\label{app:kitcomparisons}
	Table \ref{tab:kittable} compares the performance of our KIT with previous results on KITs. The pump power (third column) is the power at the KIT's input. The gain (fourth column) is the average gain observed over the KIT amplification bandwidth (fifth column). The amplitude of gain ripples over that bandwidth is indicated next to the gain value. The saturation power (sixth column) corresponds to the input $1$ dB compression point. The noise bandwidth (seventh column) is the bandwidth over which a noise measurement was performed. Finally, column eight reports the measured system noise temperature $T_\Sigma = N_\Sigma\hbar\omega / k_B $ in that bandwidth. In our case, $N_\Sigma=3.1\pm0.6$ quanta, which translates into $T_\Sigma = 0.66\pm0.15$ K. Note that Eom et.al \cite{eom2012wideband}, Ranzani et.al. \cite{ranzani2018kinetic} and Zobrist et.al. \cite{zobrist2019wide} used a wideband noise source (a hot/cold load) but did not account for the effect of the idler port's input noise. Therefore, their true system-added noise is about twice of what they reported (see appendix \ref{app:noisecorrection}). We do not compare the amplifier-added noise or amplifier-excess noise, because it is subject to approximations in the amount of loss present in the noise measurement setup. In addition, the inferred added noise usually does not include the loss of mandatory microwave components used when operating the amplifier, and as such it is an under-estimation of the true, useful amplifier-added noise.
	
	\vspace{0.1in}
	
	%
	
\end{document}